\documentclass[aps,prl,twocolumn,superscriptaddress,longbibliography]{revtex4-1}
\usepackage{braket}
\usepackage{graphicx}
\usepackage{amsmath}    
\usepackage{amsfonts}
\usepackage{lipsum}
\usepackage{xcolor}
\usepackage{qcircuit}

\begin{document}

\title{Quantum Krylov subspace algorithms \\ for ground and excited state energy estimation}

\author{Cristian L. Cortes}
\affiliation{Center for Nanoscale Materials,\\ Argonne National Laboratory, Lemont, Illinois 60439, USA}
\author{Stephen K. Gray }
\affiliation{Center for Nanoscale Materials,\\ Argonne National Laboratory, Lemont, Illinois 60439, USA}

\begin{abstract}
Quantum Krylov subspace diagonalization (QKSD) algorithms provide a low-cost alternative to the conventional quantum phase estimation algorithm for estimating the ground and excited-state energies of a quantum many-body system. While QKSD algorithms typically rely on using the Hadamard test for estimating Krylov subspace matrix elements of the form, $\langle \phi_i|e^{-i\hat{H}\tau}|\phi_j \rangle$, the associated quantum circuits require an ancilla qubit with controlled multi-qubit gates that can be quite costly for near-term quantum hardware. In this work, we show that a wide class of Hamiltonians relevant to condensed matter physics and quantum chemistry contain symmetries that can be exploited to avoid the use of the Hadamard test. We propose a multi-fidelity estimation protocol that can be used to compute such quantities showing that our approach, when combined with efficient single-fidelity estimation protocols, provides a substantial reduction in circuit depth. In addition, we develop a unified theory of quantum Krylov subspace algorithms and present three new quantum-classical algorithms for the ground and excited-state energy estimation problems, where each new algorithm provides various advantages and disadvantages in terms of total number of calls to the quantum computer, gate depth, classical complexity, and stability of the generalized eigenvalue problem within the Krylov subspace. 
\end{abstract}

\maketitle

\section*{Introduction}
The eigenpair problem for large matrices, which often consists of finding the smallest $k$ (or largest $k$) eigenvalues and eigenvectors of a matrix, remains one of the most ubiquitous problems in science. Within physics and chemistry, this problem is equivalent to finding the ground and low-lying excited state energies of a quantum many-body system represented by a large Hamiltonian matrix. While quantum computers provide a scalable route for solving this problem through the multi-ancilla-based quantum phase estimation algorithm, this approach will most likely require fault-tolerant quantum computing hardware \cite{lloyd1996universal,abrams1997simulation,zalka1998efficient,aspuru2005simulated,nielsen2002quantum,kitaev1997quantum,cleve1998quantum}. As a result, variational quantum algorithms such as the variational quantum eigensolver (VQE) \cite{peruzzo2014variational,mcclean2016theory,kandala2017hardware} and quantum approximate optimization algorithm (QAOA) \cite{farhi2014quantum} have emerged as possible candidate algorithms capable of dealing with the constraints of the current hardware in the noisy intermediate scale quantum (NISQ) era \cite{preskill2018quantum}.

Variational quantum algorithms aim to solve an optimization problem, $\text{min}_{\boldsymbol{\theta}}\; C(\boldsymbol{\theta})$, encoded through a cost function that is typically written in the form $C(\boldsymbol{\theta}) = \braket{\Psi(\boldsymbol{\theta})|H|\Psi(\boldsymbol{\theta})}$, where $H$ represents a Hermitian operator that encodes the problem of interest \cite{cerezo2021variational}. By defining a parameterized quantum circuit, $\ket{\Psi(\boldsymbol{\theta})} = U(\boldsymbol{\theta})\ket{0}^{\otimes N}$, with respect to a tunable set of parameters $\boldsymbol{\theta}$, e.g. single-qubit Pauli rotation gates, the quantum computer provides estimates of $C(\boldsymbol{\theta})$ while the classical computer performs an optimization subroutine that provides an update rule for the parameters $\boldsymbol{\theta}$ (e.g. using gradient descent). This methodology can be used to solve the eigenpair problem to estimate the ground and excited-state energies of quantum systems \cite{santagati2018witnessing,higgott2019variational,higgott2019variational,sim2018quantum}. While variational quantum algorithms have a substantial advantage in terms of gate depth, they also have significant drawbacks. 
For example, it has been shown that for a large class of quantum circuits, the optimization landscapes are highly non-convex, making the problem of finding the global minimum NP-hard \cite{bittel2021training}. It has also been shown that barren plateaus, consisting of exponentially vanishing gradients, can also arise in a wide range of conditions \cite{
mcclean2018barren,wang2020noise,cerezo2021cost,arrasmith2020effect}. For such cases, the optimization problem becomes intractable due to the inability to update the optimization parameters $\boldsymbol{\theta}$.

In recent years, quantum subspace diagonalization (QSD) methods have emerged as an alternative way of solving the eigenvalue problem for large matrices, capable of dealing with the aforementioned drawbacks \cite{huggins2020non,parrish2019quantum,seki2021quantum,stair2020multireference}. The basic idea consists of using a set of non-orthogonal quantum states, easily preparable on quantum computers, which can be used to define a generalized eigenvalue problem where the size of the corresponding matrices are exponentially smaller. The hybrid quantum-classical algorithm consists of using the quantum computer to compute the relevant matrix elements, while the generalized eigenvalue problem is solved on the classical computer (see Figure 1). The solution of the generalized eigenvalue problem provides an estimate of the relevant eigenpairs. 

A number of interesting QSD methods have been proposed which can be classified according to the numerous ways that the non-orthogonal states are defined. For instance, McClean et al. showed that by using the set of non-orthogonal basis states $a_i^\dagger a_j\ket{\Psi_G}$, it is possible to find low-lying excited states based on the ground state $\ket{\Psi_G} \approx \ket{\Psi(\boldsymbol{\theta})}$, which is found through the standard variational quantum eigensolver \cite{mcclean2017hybrid,colless2018computation}. $K$-moment states have also been proposed as an alternative way of constructing the non-orthogonal basis states, which becomes scalable when the $K$-moment unitaries are tensor products of Pauli operators \cite{bharti2020iterative,bharti2020quantum}. 

It is also possible to have provable guarantees for convergence if the set of non-orthogonal states form a Krylov basis \cite{GoluVanl96}, which is defined by the repeated application of  the 
matrix of interest, $H$, acting on the initial guess vector $\ket{\phi_o}$, resulting in the Krylov subspace $\mathcal{K}_M = \text{span}\{\ket{\phi_o},H\ket{\phi_o},H^2\ket{\phi_o},\cdots,H^{M-1}\ket{\phi_o} \}$. The Lanczos method \cite{Parlett1980} is one of the most well-known algorithms that uses this subspace to solve the eigenpair problem with convergence properties that are dependent on the spectral properties of the matrix $H$ as well as the overlap of the guess vector $\ket{\phi_o}$ with the true solution. While this method is routinely executed on classical computers, its implementation on a quantum computer is more challenging since $H$ is not unitary. Nonetheless, interesting approaches that invoke sums of unitary operators as approximations to the Hamiltonian matrix and its higher powers have recently been suggested \cite{seki2021quantum,  bespalova2021hamiltonian} and remains an ongoing research direction. Motta et al. have also proposed the QLanczos algorithm  which defines the Krylov subspace by the repeated application of the imaginary time evolution propagator, $f(\hat{H}) = e^{-\beta\hat{H}}$ \cite{motta2020determining}. In this framework, the non-unitary imaginary-time propagator is written as a unitary operator under the condition that the Hamiltonian is $k$-local. A linear system of equations must be solved classically for each imaginary time step, where the number of measurements and size of such equations grows exponentially with the spreading of entanglement. 

\begin{figure}[t!]
    \centering
    \includegraphics[width=8.5cm]{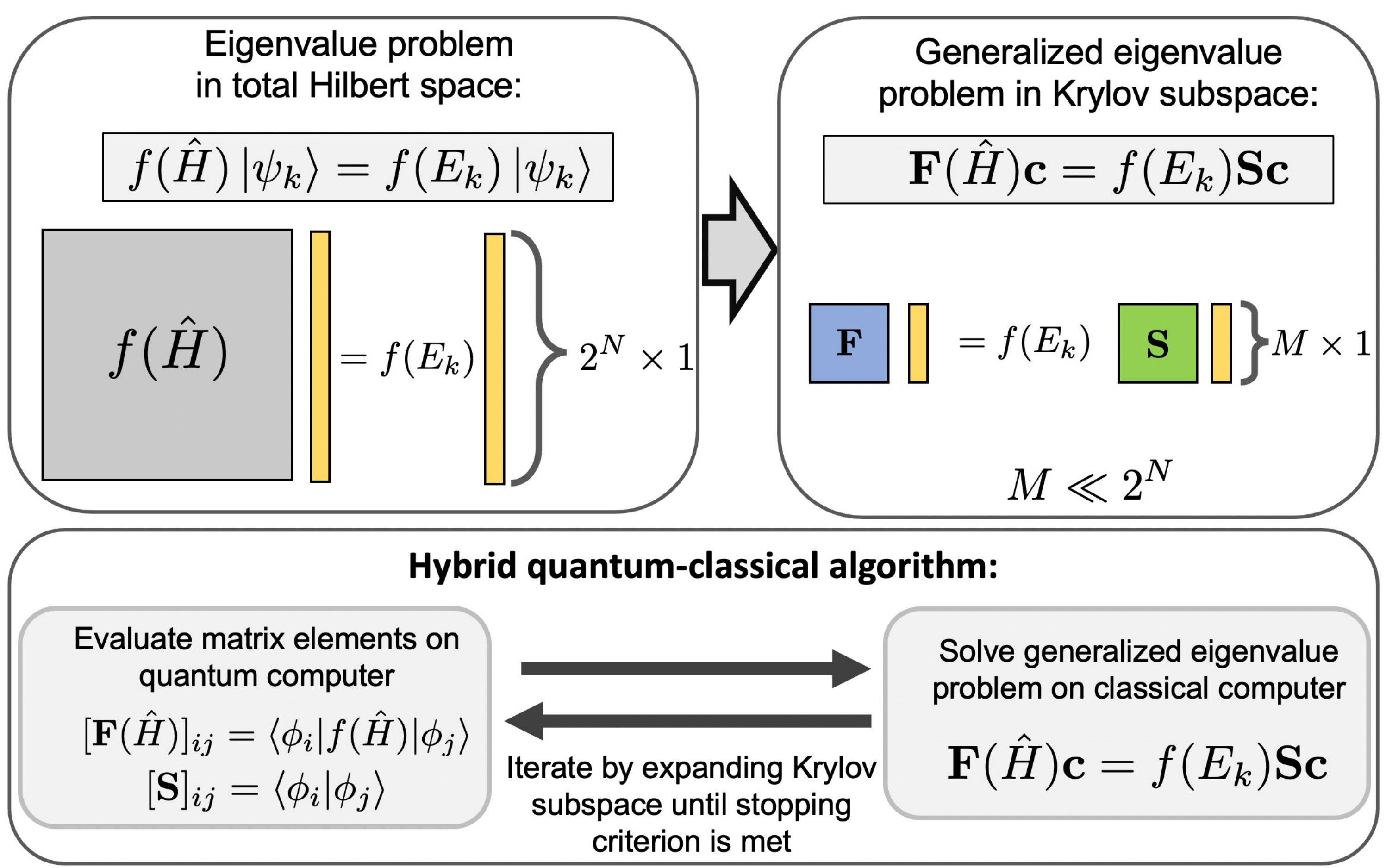}
    \caption{Overview of quantum Krylov subspace algorithms.}
    \label{fig:my_label}
\end{figure}

In this manuscript, we focus on solving the eigenpair problem with sets of Krylov basis states generated by real-time quantum dynamics.  This idea, in the context of quantum computing approaches to the eigenpair problem, was pioneered by Parrish and McMahon \cite{parrish2019quantum}. They referred to their approach as the quantum filter diagonalization (QFD) algorithm, because of similarities with the classical-computer-based filter diagonalization methods developed in the  1990s \cite{wall1995extraction,mandelshtam1997low,mandelshtam2001fdm}.  Independently, Stair et al. proposed a  multi-reference selected quantum Krylov subspace (MRSQK) algorithm \cite{stair2020multireference} which can be viewed as a generalization of QFD. These methods represent variants of the QLanczos algorithm where the real-time evolution operator,  $e^{-i\hat{H}\tau}$, is used to generate the Krylov basis, where $\tau$ is equal to the time step size and we assume atomic units such that $\hbar$ = 1 throughout this manuscript. While these methods have shown great promise, they are not without practical issues with respect to NISQ-era applications. First, quantum Krylov subspace algorithms based on real-time dynamics require Hadamard test quantum circuits (see Appendix for more details), which uses an ancilla qubit with controlled multi-qubit controlled unitary operations \cite{parrish2019quantum,stair2020multireference}. This approach substantially increases the circuit depth, making it more difficult for NISQ-era hardware. Second, the number of calls to the quantum computer that are required to construct the Krylov subspace matrices scales as $\mathcal{O}(LM^2)$ where $L$ is the number of terms in the Hamiltonian and $M$ is the subspace matrix size \cite{stair2020multireference}. Third, single-reference Krylov subspace algorithms also suffer from large condition numbers (ratio of largest to smallest singular values) of the overlap matrix $\mathbf{S}$ that become substantially worse as the number of time steps increases. In principle, this could make the solution of the generalized eigenvalue problem not possible for many problems of interest, such as strongly correlated systems.

In this work, we provide several major contributions which address the outstanding problems discussed above. First, we show that the Hadamard test is not required to estimate the Krylov subspace elements of the form $\braket{\phi_i|e^{-i\tau\hat{H}}|\phi_j}$ for a large class of Hamiltonians relevant to nuclear physics, quantum chemistry, and condensed matter physics. Our approach avoids the need for an ancilla qubit with controlled unitary operations and, when combined with efficient fidelity estimation protocols, provides a substantial reduction in circuit depth compared to previous approaches.  Our second major contribution includes the proposal of three new generalized eigenvalue problems which can be used to estimate both ground and excited-state energies. 
Each of these generalized eigenvalue problems provides various advantages and disadvantages in terms of the total number of calls to the quantum computer, gate depth, classical post-processing complexity, as well as stability of the generalized eigenvalue problem based on the condition number of the overlap matrix $\mathbf{S}$. 

In particular, we show that two of the newly proposed generalized eigenvalue problems involving the unitary function $e^{-i\hat{H}\tau}$ only require $\mathcal{O}(M)$ calls to the quantum computer compared to the $\mathcal{O}(LM^2)$ calls that are required for QKSD algorithms which use Hamiltonian-based generalized eigenvalue problems. We also  show that two of the newly proposed generalized eigenvalue problems, which also use real-time quantum dynamics to generate the Krylov space as in the QFD approach \cite{parrish2019quantum} more  closely resemble the original classical filter diagonalization method (FDM) originally proposed by Wall and Neuhauser  \cite{wall1995extraction} and subsequently further developed and elaborated upon by Mandelshtam and Taylor \cite{mandelshtam1997low} in that they target specific energy ranges by energy filtering. Compared to typical Krylov subspace algorithms which have a classical computational complexity scaling that is polynomial in the total number of time steps, the FDM method provides a constant time scaling $\mathcal{O}(1)$. We also find empirical evidence that the corresponding generalized eigenvalue problems have condition numbers that are orders of magnitude smaller than the single-reference Krylov subspace counterparts as the number of time steps increases. To test the efficacy of the proposed algorithms, we numerically compare these four methods for the problem of finding the ground-state and excited-state energies of various quantum chemistry Hamiltonians, showing fast convergence with a small number of discrete time steps.

It is worth noting that while we were finalizing this manuscript, we became aware of the paper by Klymko \emph{et al.} \cite{klymko2021real}, currently in preprint, which is similar in spirit to our work. Klymko \emph{et al.} provide a theoretical basis for non-orthogonal states generated by real-time dynamics, and independently proposed a hybrid quantum-classical algorithm based on the generalized eigenvalue problem with the unitary function, $e^{-i\hat{H}\tau}$, which they refer to as variational QPE or VQPE. This method is equivalent to one of the three new methods that we present in this manuscript, which we refer to as KDM U. In addition, their major contributions include a comprehensive study of the effects of noise, Trotter-Suzuki error, and a comparison between VQPE and the conventional QPE algorithm in terms of the total simulation time and total number of time steps required to reach chemical accuracy. 
Our major contributions include the multi-fidelity estimation protocol which may be used to avoid the Hadamard test, as well as the two hybrid quantum-classical algorithms based on Fourier filter energies, which more closely resemble the classical filter diagonalization method of Wall and Neuhauser, and Mandelshtam and Taylor \cite{wall1995extraction,mandelshtam1997low}, which we refer to as the FDM H and FDM U methods.  

\section{Quantum Krylov subspace diagonalization method}
We aim to find the ground and excited-state energies of a general many-body Hamiltonian written as a sum of $N$-qubit Pauli terms, 
$\hat{H} = \sum_i^L h_i \hat{P}_i$, where $h_i$ is a weighting coefficient and $\hat{P}_i$ is a general tensor product of $N$ Pauli operators, ${\hat{P}}_i = \otimes_{k=1}^{N_i} \hat{\sigma}_{i_k}^{(\mu_k)}$, with $\mu_k$ denoting the qubit number and $i_k$ acts as a label for the type of Pauli operator $\{\hat{I}, \hat{\sigma}_x,\hat{\sigma}_y,\hat{\sigma}_z\}$.  We do not impose any type of restrictions on the locality of the Hamiltonian, thereby allowing for the implementation of the proposed algorithms for many problems relevant to nuclear physics, condensed matter physics, and quantum chemistry. For illustration purposes, we will focus on the canonical quantum chemistry Hamiltonian in second quantization which is able represent the electronic structure problem of a wide variety of molecular systems (see appendix for details). 

The Hamiltonian $\hat{H}$ obeys the standard eigenvalue equation, $\hat{H}\ket{\psi_k} = E_k \ket{\psi_k}$, with the energy eigenvalue $E_k$ and corresponding eigenvector $\ket{\psi_k}$, assumed to satisfy the orthonormality condition,  $\braket{\psi_{k'}|\psi_k}=\delta_{kk'}$. General functions of the Hamiltonian $f(\hat{H})$ will also obey the eigenvalue equation,
\begin{equation}
    f(\hat{H})\ket{\psi_k} = f(E_k)\ket{\psi_k}.
    \label{Eigenvalue}
\end{equation}
The matrix size for this eigenvalue problem scales exponentially with the total number of qubits.  However, as we show below, this equation can be used to define a wide variety of generalized eigenvalue problems in a subspace that is exponentially smaller. The basic idea requires expanding the eigenvector $\ket{\psi_k}$ as a linear combination of non-orthogonal states $\ket{\phi_n}$,
\begin{equation}
    \ket{\psi_k} \approx \sum_{n=0}^{M-1} c_n \ket{\phi_n}.
\end{equation}
Substituting this result into (\ref{Eigenvalue}) and multiplying from the left by $\bra{\phi_{n'}}$, we find the generalized eigenvalue problem,
\begin{equation}
    \mathbf{F}(\hat{H})\mathbf{c} = f(E_k)
    \mathbf{S}\mathbf{c},
    \label{GeneralizedEigenvalueEquation}
\end{equation}
where $\mathbf{c} = (c_0,c_1,\cdots,c_{M-1})^T$ is a column vector of expansion coefficients and the subspace matrices $\mathbf{F}(\hat{H})$ and $\mathbf{S}$ are defined by the matrix elements,
\begin{align}
    [\mathbf{F}(\hat{H})]_{nn'} = \braket{\phi_n|f(\hat{H})|\phi_{n'}}  \;\text{and}\;
    [\mathbf{S}]_{nn'} = \braket{\phi_n|\phi_{n'}}.
    \label{overlap}
\end{align}
Naturally, the subspace matrix size is much smaller in the non-orthogonal basis when $M\ll 2^N$, and is also more general than Hamiltonian-based generalized eigenvalue problems derived in previous works. The choice of Hamiltonian function $f(\hat{H})$ and non-orthogonal basis, $\ket{\phi_n}$, will ultimately lead to a wide variety of different hybrid quantum-classical algorithms with trade-offs in terms of convergence, number of calls to the quantum computer, circuit depth, and classical complexity required for post-processing. Although it might not be implementable in the near-term, it is worth mentioning that a non-orthogonal basis defined by the function, $f(\hat{H}) = \exp\left( i\arccos{(\hat{H}\tau)}\right)$, as used in qubitization \cite{low2019hamiltonian}, would provide an effective way to estimate both ground and excited-state energies with equivalent circuit depths as a single Trotter time step but avoiding Trotter error \cite{poulin2018quantum}. However, this comes at a cost adding a register of ancilla qubits with controlled multi-qubit unitary gates. For our purposes, we will focus on the standard Hamiltonian and real-time evolution operators, $f(\hat{H}) = \hat{H}$ and $f(\hat{H}) = e^{-i\hat{H}\tau}$. We will also consider two different sets of non-orthogonal states, thereby obtaining four different quantum-classical algorithms which provide various advantages and disadvantages as discussed below.


\paragraph{Krylov subspace diagonalization method (KDM).} The Krylov subspace diagonalization method assumes the eigenvector $\ket{\psi_k}$ may be written as a linear combination of real-time evolved Krylov basis states,
\begin{equation}
    \ket{\psi_k} \approx \ket{\psi_K} =  \sum_{n=0}^{M-1} c_n e^{-in\hat{H}\tau}\ket{\phi_o} =  \sum_{n=0}^{M-1} c_n \ket{\phi_n}
\end{equation}
where $\ket{\phi_o}$ is the initial single-reference state. Using the steps outlined above, the corresponding generalized eigenvalue problem may be written as, $\mathbf{F}_K(\hat{H})\mathbf{c}_K = f(E_k)\mathbf{S}_K\mathbf{c}_K$, where the subscript $K$ denotes the real-time Krylov basis with the subspace matrix elements defined using equation (\ref{overlap}). We emphasize that this basis, specifically when $f(\hat{H}) = \hat{H}$, which we refer to as the KDM H method, recovers the QFD and MRSQK methods in the limit of a single reference state \cite{parrish2019quantum,stair2020multireference}. Our new contribution therefore corresponds to the KDM method with $f(\hat{H}) = e^{-i\hat{H}\tau}$, which we refer to as the KDM U method, which also coincides with the recent proposal by Klymko et al. \cite{klymko2021real}. 

\paragraph{Filter Diagonalization Method (FDM).} The filter diagonalization method, on the other hand,  approximates $\ket{\psi_k}$ as a linear combination of time-evolved wavefunctions that are Fourier transformed with respect to a set of filter energies $E_j$,
\begin{align}
    \ket{\psi_k}\approx \ket{\psi_F} = \sum_{j=1}^J \sum_{n=0}^{M-1} c_j   e^{-in(\hat{H}-E_j)\tau}\ket{\phi_o} = \sum_j^J c_j \ket{\phi_j}
\end{align}
resulting in the generalized eigenvalue problem, $\mathbf{F}_J(\hat{H})\mathbf{c}_J = f(E_k)\mathbf{S}_J\mathbf{c}_J$ with matrix elements defined by the non-orthogonal basis of filter energies. This type of basis has an interesting property in that $\ket{\phi_j}$ will be dominated by eigenvectors whose eigenvalues are close to the filter energies $E_j$. If we expand the starting state, $\ket{\phi_o}$, in terms of the true eigenvectors $\ket{\psi_{k'}}$ of the Hamiltonian, such that $\ket{\phi_o} = \sum_{k'} c_{k'} \ket{\psi_{k'}}$, then it is possible to show that,
\begin{equation}
    \ket{\phi_j} = \sum_{k'} c_{k'} \frac{1-e^{-iM(E_{k'}-E_j)\tau}}{1-e^{-i(E_{k'}-E_j)\tau}}\ket{\psi_{k'}},
\end{equation}
highlighting the fact that eigenvectors with eigenvalues $E_{k'}$ that are close to the filter energies $E_j$ ($E_{k'}\sim E_j$) will have the largest contribution. In total, the FDM method will give rise to two new quantum-classical algorithms which we refer to as the FDM H method if $f(\hat{H})= \hat{H}$ and FDM U method if $f(\hat{H}) = e^{-i\hat{H}\tau}$.

\paragraph{Relationship between both non-orthogonal bases.}
Analyzing the two generalized eigenvalue problems, it is possible to show that the two approaches are related by the $M\times J$ transformation matrix,
\begin{equation*}
    \mathbf{W} = 
    \begin{pmatrix}
    1 & 1 & \cdots & 1 \\
    e^{-iE_1\tau} & e^{-iE_2\tau} & \cdots & e^{-iE_J\tau} \\
    e^{-i2E_1\tau} & e^{-i2E_2\tau} & \cdots & e^{-i2E_J\tau} \\
    \vdots & \vdots & \ddots & \vdots \\
    e^{-i(M-1)E_1\tau} & e^{-i(M-1)E_2\tau} & \cdots & e^{-i(M-1)E_J\tau} \\
    \end{pmatrix}
\end{equation*}
resulting in the following relationship between the real-time evolution Krylov diagonalization method and the filter diagonalization method,
\begin{align}
    \mathbf{W}^\dagger \mathbf{F}_K(\hat{H}) \mathbf{W} &= \mathbf{F}_J(\hat{H}) \label{relation1}\\
    \mathbf{W}^\dagger \mathbf{S}_K \mathbf{W} &= \mathbf{S}_J.
    \label{relation2}
\end{align}
In the case that the filter frequencies are chosen with an equidistant grid, such that $E_j = \frac{2\pi}{M\tau} j$ where $j=0,\cdots,M-1$, then the transformation $\mathbf{W}$ becomes a unitary matrix up to a normalization factor equivalent to the discrete Fourier transform matrix. It is important to emphasize, however, that the total number of discrete energies $J$ can be much smaller than the total number of time steps, $M$, resulting in a constant-time $\mathcal{O}(1)$ computational complexity for solving the FDM-based generalized eigenvalue problem on the classical computer, compared to the polynomial scaling $\mathcal{O}(\text{poly}(M))$ for the KDM method. Moreover, the choice of filter window with a suitable number of filter energies can also stabilize the generalized eigenvalue problem resulting in smaller condition numbers for the overlap matrix $\mathbf{S}$. We provide evidence of this result in the numerical experiments section of the manuscript. 

\begin{table*}
  \centering
    \begin{tabular}{ |c|c|c|c|  }
     \hline
    $f(\hat{H})$ & Non-orthogonal Basis & $N_K$ & Classical Complexity \\
     \hline
     $\hat{H}$   & real-time dynamics & $\mathcal{O}(LM^p/\epsilon^2)$ & $\mathcal{O}(\text{poly}(M))$ \\
     $e^{-i\hat{H}\tau}$ & real-time dynamics & $\mathcal{O}(M/\epsilon^2)$ & $\mathcal{O}(\text{poly}(M))$ \\
     $\hat{H}$   & Fourier filter energies  & $\mathcal{O}(LM^p/\epsilon^2)$ & $\mathcal{O}(1)$ \\
     $e^{-i\hat{H}\tau}$ &  Fourier filter energies  & $\mathcal{O}(M/\epsilon^2)$ & $\mathcal{O}(1)$ \\
     \hline
    \end{tabular}
  \caption{Summary of results. $N_K$ is equal to the number of measurements/calls to the quantum computer required to construct the Krylov subspace matrix elements. $L$ is equal to the total number of terms in the Hamiltonian $\hat{H}$. $M$ is equal to the order of real-time evolved Krylov subspace. The exponent $p$ is equal to one if the Hamiltonian, $\hat{H}$, perfectly commutes with the quantum circuit unitary which approximates the time-evolution operator, $e^{-i\tau\hat{H}}$, otherwise it is equal to two.}
  \label{tab:1}
\end{table*}

\paragraph{Number of calls to the quantum computer.} In the following, we provide an estimate of the number of calls to the quantum computer, $N_K$, required to construct the Krylov subspace matrix elements defined by Equation (\ref{overlap}). We assume that the subspace matrix elements are computed with Hadamard-test quantum circuits, or equivalently, with the multi-fidelity estimation protocol which we describe in the following section. For the latter, our estimates are based on single fidelity estimation circuits, such as a \texttt{SWAP}-test circuit or a {mirror}-like quantum circuit, as outlined in the Discussion section below. In general, the estimation of these quantities to precision $\epsilon$ will require $1/\epsilon^2$ samples, which will result in a $1/\epsilon^2$ multiplicative factor for all of the cases we consider below. Furthermore, we will restrict ourselves to the single-reference Krylov subspace algorithm, though more general estimates of the multi-reference case may be done with the same arguments. 

We first consider the estimation of the overlap matrix elements in $\mathbf{S}$, noting that they will be the same for all four methods. These matrix elements are equivalent to correlation functions of the form $C_n(\tau) = \braket{\phi_o|e^{-in\hat{H}\tau}|\phi_o}$. By assuming that a single call to the quantum computer provides an estimate of both the real and imaginary parts of the correlation function $C_n(\tau)$, then $M-1$ calls to the quantum computer are required to construct the overlap matrix $\mathbf{S}$.

For $f(\hat{H})= \hat{H}$, the matrix elements of the subspace matrix $\mathbf{F}(\hat{H})$ may be written as $\braket{\phi_n|\hat{H}|\phi_{n'}} = \braket{\phi_o|e^{in\hat{H}\tau}\hat{H}e^{-in'\hat{H}\tau}|\phi_o}$. Here, the number of calls to the quantum computer will depend on whether the Hamiltonian $\hat{H}$ and the quantum circuit implementation of the time-evolution operator, $e^{-in\hat{H}\tau}$, commute. If we assume that they commute, these elements may be written as $\braket{\phi_o|\hat{H}e^{-i(n'-n)\hat{H}\tau}|\phi_o}= \sum_i^L h_i \braket{\phi_o|\hat{P}_ie^{-i(n'-n)\hat{H}\tau}|\phi_{o}}$, resulting in a Toeplitz matrix structure that requires $\mathcal{O}(LM)$ calls to the quantum computer. However, in the case of Trotterized quantum circuits where the commutation relation does not hold exactly, the total number of calls would scale as $\mathcal{O}(LM^2)$. 

Methods using the real-time evolution function, $f(\hat{H}) = e^{-i\hat{H}\tau}$, will have a complexity that is substantially less. In this case, the matrix elements will also correspond to correlation functions, $C_n(\tau)$. The matrix elements from the overlap matrix, $\mathbf{S}$, can therefore be used to construct the majority of the matrix elements in $\mathbf{F}(\hat{H})$. The off-diagonal elements in the top-right and bottom-left corners, however, will require an additional call to the quantum computer for the estimation of the $C_{M}(\tau)$ correlation function. In total, the $f(\hat{H}) = e^{-i\hat{H}\tau}$ method will require $M$ calls to the quantum computer. This, however, comes at the cost of requiring a single quantum circuit evaluation with an increased circuit depth equivalent to a single time step (assuming a Trotterized time-evolution circuit).

Finally, it is worth noting that for fixed $f(\hat{H})$, the KDM and FDM methods will have an equivalent number of calls to the quantum computer because they
are related by equations (\ref{relation1}) and (\ref{relation2}). This highlights the fact that both methods only differ in the post-processing methodology used to estimate the ground and excited-state energies and, as a result, both methods can be carried out in parallel on a classical computer. A summary of these results is shown in Table I, underlining how each of these four methods carries different complexities due to quantum and classical computational resources. Here, we wrote the scaling of the number of calls $N_K$ for $f(\hat{H}) = \hat{H}$ as $\mathcal{O}(L M^p)$ where $p=1$ if the Hamiltonian and time-evolution unitary circuit commutes and $p=2$ otherwise.  

\section{Multi-fidelity estimation protocol}
We now consider the evaluation of the complex-valued matrix elements (\ref{overlap}), equivalent to a single call the quantum computer as defined above. The Hadamard test is the standard method used for estimating these types of non-Hermitian quantities (see Appendix D for more details), which originates from the single-ancilla-based quantum phase estimation algorithm from Kitaev. This approach requires an ancilla qubit with controlled unitary operations that substantially increases the total number of multi-qubit gates in the overall circuit. In the near term, multi-qubit gates (e.g. CNOT gates) represent an expensive resource. In the following, we propose a method that avoids the Hadamard test thereby making a wide variety of quantum Krylov subspace diagonalization methods more amenable to near-term devices. 

The motivation for our proposed method stems from the field of interferometry which aims to measure a target phase $\theta_t$ that encodes a physical parameter of interest. Interference pattern measurements can only provide information about the phase difference, $\Delta\theta = \theta_r-\theta_t$. A reference laser is typically used to provide a controllable reference phase $\theta_r$, allowing for the proper estimation of $\theta_t$. While the Hadamard test provides a reference phase through use of the ancilla qubit, the multi-fidelity estimation (MFE) protocol generates the reference phase through the superposition state $\tfrac{1}{\sqrt{2}}(\ket{R} + \ket{\phi_k})$, where the reference state $\ket{R}$ and the target state $\ket{\phi_k}$ originate from different symmetry sectors of the Hamiltonian. If the time evolution of the reference state is classically simulatable, it will be possible to have a reference phase without an ancilla qubit. Below, we provide a more detailed mathematical description.

The proposed approach assumes that the Hamiltonian contains a symmetry $\hat{S}$ such that, $[\hat{H},\hat{S}] = 0$, with quantum states $\ket{\phi_k}$ that have a definite symmetry, such that $\braket{\phi_k|\hat{S}|\phi_k} = s_k$, where $s_k$ corresponds to an eigenvalue of the symmetry operator $\hat{S}$. We also assume that there exists a reference state $\ket{R}$ where $\braket{R|e^{-in\tau\hat{H}}|R}$ is efficient to calculate on the classical computer.  We emphasize that the reference and target states, $\ket{R}$ and $\ket{\phi_k}$, are not required to be eigenstates of the Hamiltonian, but they do need to originate from different symmetry sectors such that $\braket{R|\phi_k} = 0$. Note that this condition relaxes the requirements from conventional quantum phase estimation algorithms that normally require the initial state to be an eigenstate of the Hamiltonian. If all of these conditions hold, then it will be possible to implement the multi-fidelity estimation protocol as shown below. For many nuclear physics, quantum chemistry and condensed matter physics applications, the particle number, total spin and spin projection symmetries may be applicable and can be used in this approach due to the fact that they contain symmetry sectors that are classically tractable. 

As a concrete example, we consider the quantum chemistry Hamiltonian which conserves the electron number $\hat{N} = \sum_i \hat{a}_i^\dagger \hat{a}_i$. We assume that the quantum states $\ket{\phi_k}$ have a definite electron number, $\braket{\phi_k|\hat{N}|\phi_k} = n_k$, that is not equal to zero. To estimate the most general off-diagonal element, $\braket{\phi_i|e^{-in \hat{H}\tau}|\phi_j}$, as required by multi-reference Krylov-based methods, the MFE protocol requires measuring the following state fidelities on the quantum computer,
\begin{align}
    F_{1} &= |\braket{\phi_i|e^{-in \hat{H}\tau}|\phi_j}|^2,  \label{F1}\\
    F_{2} &= \tfrac{1}{4}|(\bra{\phi_i} + \bra{R})|e^{-in \hat{H}\tau}|(\ket{R}+\ket{\phi_j})|^2  \label{F2}.
\end{align}
Combining both results yields the magnitude and phase of the off-diagonal matrix element $\braket{\phi_i|e^{-in \hat{H}\tau}|\phi_j} = re^{i\theta}$, written in complex polar coordinates, 
\begin{align}
    r    &= \sqrt{F_1} \\
    \theta &= \cos^{-1}\left(\frac{4F_2-F_1-r_R^2}{2r_R\sqrt{F_1}} \right) + \theta_R
\end{align}
where $r_R$ and $\theta_R$ represent the reference amplitude and phase defined as, $\braket{R|e^{-in\hat{H}\tau}|R} = r_Re^{i\theta_R}$. If we take the reference state as the zero particle number (vacuum) state, $\ket{R}\equiv\ket{0}^{\otimes N}$, then the reference amplitude $r_R$ will be equal to one while the reference phase will be equal to $\theta_R = -n\tau\braket{0|\hat{H}|0}$, where $\braket{0|\hat{H}|0}$ denotes the expectation value of the Hamiltonian with respect to the vacuum state which can be evaluated efficiently on a classical computer. 

\subsection{Preparation of $\tfrac{1}{\sqrt{2}}(\ket{R} + \ket{\phi_k})$}
One of the key requirements of this protocol is the preparation of the state $\tfrac{1}{\sqrt{2}}(\ket{R}+ \ket{\phi_k})$ on the quantum computer. Note that because we have imposed the requirement that $\ket{R}$ and $\ket{\phi_k}$ belong to different symmetry sectors (i.e. contain different particle numbers), it is possible to prepare such states using $\texttt{GHZ}$-state-preparation circuits. For instance, we consider the preparation of the state $\tfrac{1}{\sqrt{2}}(\ket{R} + \ket{\phi_{\text{HF}}})$ where $\ket{R}$ is the vacuum state and  $\ket{\phi_{\text{HF}}}$ is the Hartree-Fock state for a system of $N$ spin-orbitals (represented by $N$ qubits) and $\eta$ electrons. In the Jordan-Wigner basis, the Hartree-Fock state takes the simple product-state form, $\ket{\phi_{\text{HF}}} = \ket{0}^{\otimes N-\eta}\otimes\ket{1}^{\otimes\eta}$, where the first $\eta$ qubits are prepared in the one state and the rest of the qubits remain in the zero state. To prepare the target superposition  state, $\tfrac{1}{\sqrt{2}}(\ket{0}^{\otimes N}+ \ket{\phi_{\text{HF}}})$, a Hadamard gate is applied to the first qubit, followed by a ladder of CNOT gates applied up to the $\eta$th qubit, resulting in a total of $\eta-1$ CNOT gate operations. More general states can also be prepared by subsequently applying a symmetry-conserving quantum circuit $U_S$ \cite{gard2020efficient}, which could in principle represent a parameterized quantum circuit originating from a VQE pre-processing step. As an example, we provide the quantum circuit that prepares the state, $\tfrac{1}{\sqrt{2}}(\ket{000000} + \ket{000111})$, where the Hartree-Fock state represents a system of six spin-orbitals with three electrons.
\[
\Qcircuit @C=2em @R=1em @!R 
{
\lstick{\ket{0}}  & \gate{H} & \ctrl{1} &  \qw & \multigate{5}{U_S} & \qw \\
\lstick{\ket{0}}  & \qw &  \targ & \ctrl{1} & \ghost{U_S} & \qw \\
\lstick{\ket{0}}   & \qw &   \qw & \targ  & \ghost{U_S} & \qw \\
\lstick{\ket{0}}   & \qw & \qw & \qw & \ghost{U_S} & \qw \\
\lstick{\ket{0}}   & \qw & \qw & \qw & \ghost{U_S} & \qw \\
\lstick{\ket{0}}   & \qw & \qw & \qw & \ghost{U_S} & \qw \\
&&\mbox{$\tfrac{1}{\sqrt{2}}(\ket{00000} + \ket{000111})$} 
\gategroup{1}{2}{6}{4}{1.5em}{--}
}
\]
\newline

\begin{figure*}[t!]
    \centering
    \includegraphics[width=18cm]{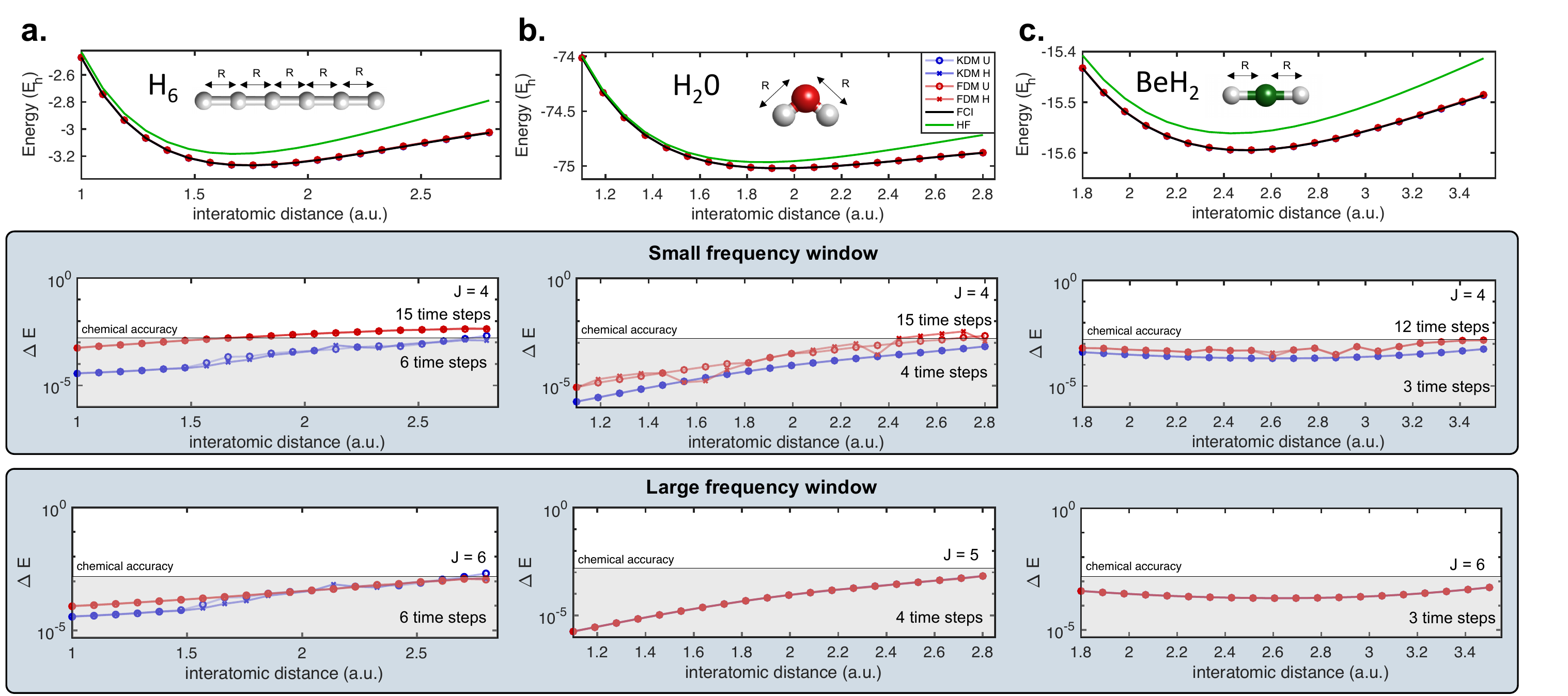}
    \caption{Comparison of KDM and FDM methods for predicting the ground-state potential energy curves of a (a) 6-site linear hydrogen chain, (b) H$_2$O molecule, and (c) BeH$_2$ molecule. The top row plots the absolute energy scale measured in Hartrees. The second and third rows plot the energy error $\Delta E = |E_{\text{FCI}}-E_{\text{approx}}|$ between the full configuration interaction calculation and the proposed approximate methods, where the second row uses a narrow frequency window, $[E_{\text{HF}}- 0.3 E_h,E_{\text{HF}}+ 0.2 E_h]$ for the FDM method, while the third row uses a large frequency window, $[E_{\text{HF}}- 20 E_h,E_{\text{HF}}+ 20 E_h]$.}
    \label{fig:Figure2}
\end{figure*}

\section*{Numerical Experiments}

In Figure 2, we compare these four methods for the estimation of the ground-state energy curves for three different molecular systems: (a) a one-dimensional $H_6$ chain, (b) an H$_2$O molecule with fixed bond angle $\phi=104.45^\circ$, and (c) a BeH$_2$ molecule. The energy curves are plotted as a function of interatomic distance $R$ labelled in the insets. See Appendix A for a discussion of the various levels of electronic structure theory used in these examples. In all cases, we numerically simulate the hybrid quantum-classical algorithm as described in Figure 1 using the Hartree-Fock (HF) state as the initial single-reference state $\ket{\phi_o}$ and an ideal time-evolution quantum circuit $U(n\tau) = e^{-in\hat{H}\tau}$ where the time step size, $\tau = 0.1$ a.u., is used for all of the simulations. The choice of time step is very important but can be a subtle issue that is discussed in Appendix B.  In the top row, we plot the absolute energy scale measured in Hartrees while the bottom two rows show the energy error $\Delta E = |E_{\text{FCI}} - E_{\text{approx}}|$. We also compare the FDM method with a small frequency window $[E_{\text{HF}}- 0.3 E_h,E_{\text{HF}}+ 0.2 E_h]$ in the second row and a large frequency window $[E_{\text{HF}}- 20 E_h,E_{\text{HF}}+ 20 E_h]$ in the third row. The conventional implementation of the FDM method uses a narrow frequency window to select the filter energies $E_j$ \cite{mandelshtam1997low}, corresponding to the methodology used in the second row. Using this approach, we find that the FDM method converges much more slowly than the equivalent KDM method, thereby requiring more time steps to achieve similar accuracy. Instead we found that by using large frequency windows to choose the filter energies $E_j$, it is possible to improve the convergence rate of the FDM method so that it is comparable to the KDM method. This is shown by the nearly overlapping energy error curves in the third row. In the latter case, we find that we are able to achieve chemical accuracy within 3 to 6 time steps for all three molecular systems. While we do not consider Trotter error in this work, it is worth noting that previous work \cite{parrish2019quantum,stair2020multireference} has shown that Trotterized quantum circuits provide an additional error, $\Delta_t$, that increases the energy error $\Delta E$ of the ground-state energy estimate. We expect similar behavior for the quantum-classical algorithms that we have proposed in this manuscript, but will leave such studies for future work. 

Figure 2 demonstrates the dependence of the energy error of the FDM method as a function of the hyperparameters:  $E_{\text{min}}$, $E_{\text{max}}$, and $J$ which define the energy window $[E_\text{min},E_\text{max}]$ as well as the total number of filter energies, $J$. The choice of energy window and number of filter energies affects the convergence rate of the FDM method quite dramatically, which is also illustrated clearly in Figure 3. In practice, a hyperparameter optimization loop can be added to the conventional hybrid quantum-classical algorithm depicted in Figure 1, which would be performed  on the classical computer exclusively without any additional calls to the quantum computer. For instance, this loop may consist of an exhaustive grid search of the parameters $E_{\text{min}}$, $E_{\text{max}}$, and $J$ while monitoring either the ground-state energy estimate or the variance of the function $f(\hat{H})$ as a function of the FDM wavefunction, $\ket{\psi_F} = \sum_j^J c_j \ket{\phi_j}$. For the ground-state energy estimation problem, $c_j$ would correspond to the eigenvector coefficients from the generalized eigenvalue equation (\ref{GeneralizedEigenvalueEquation}) with the smallest eigenvalue. Explicitly, the hyperparameter optimization loop for the FDM U method would monitor the variance of the function, $f(\hat{H}) = e^{-i\hat{H}\tau}$, written as
\begin{equation}
    \text{Var}[e^{-i\hat{H}\tau}] = 1 - |\!\!\braket{\psi_F|e^{-i\hat{H}\tau}|\psi_F}\!\!|^2,
    \label{variance}
\end{equation}
where we have assumed that the function $\ket{\psi_F}$ has been normalized, such that $\braket{\psi_F|\psi_F} = 1$. This definition of the variance has a maximum value of 1 and minimum value of zero when $\ket{\psi_F}$ is an eigenstate of the Hamiltonian, $\hat{H}$. It is important to note that the second term on the right hand side of equation (\ref{variance}) may be written as a sum of the matrix elements defined by equation (4), $\braket{\phi_n|e^{-i\hat{H}\tau}|\phi_{n'}}$, highlighting the fact that the variance is easily calculated on the classical computer with no additional calls to the quantum computer. 

\begin{figure}[t!]
    \centering
    \includegraphics[width=8.7cm]{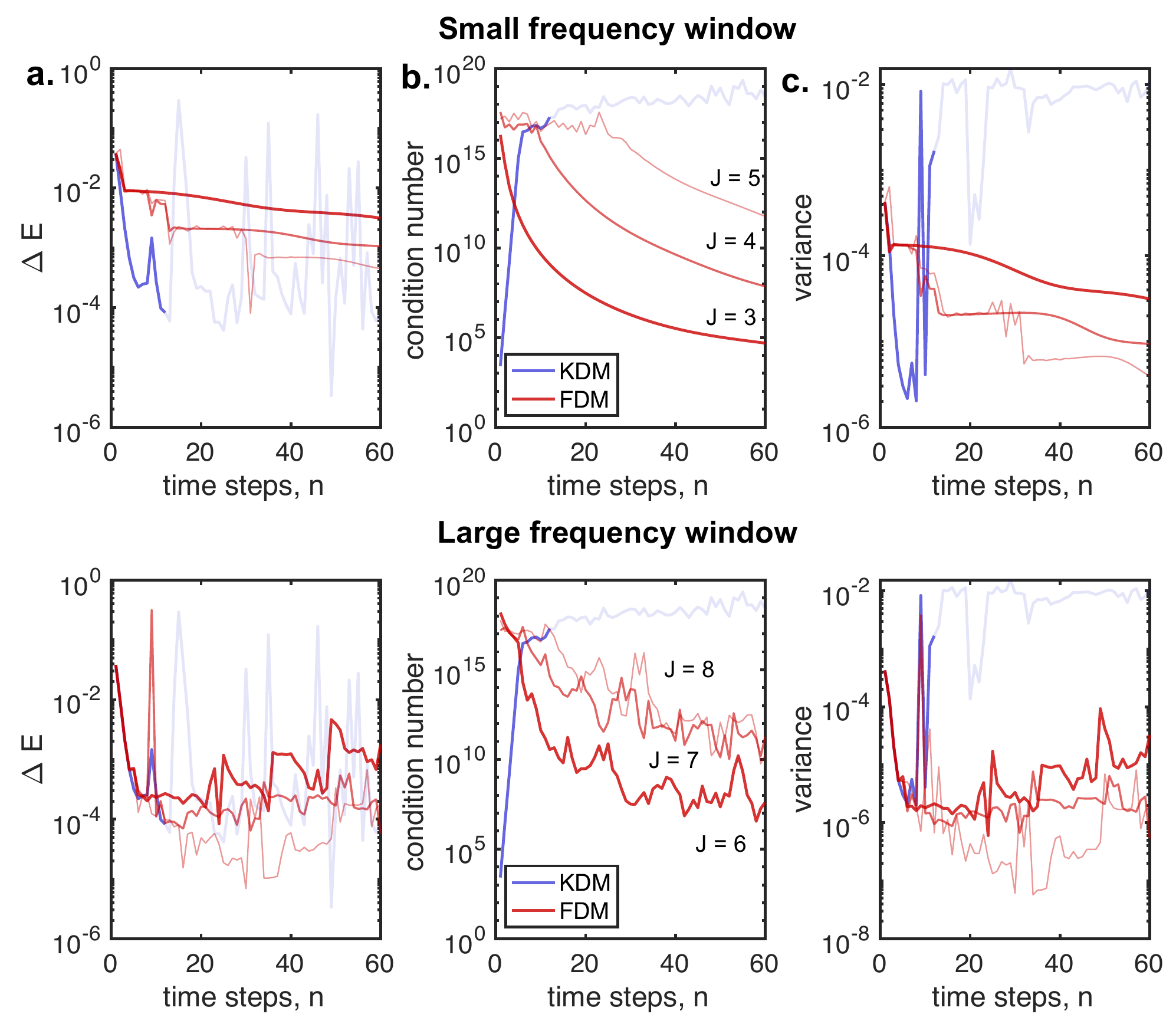}
    \caption{(a) Energy error $\Delta E$, (b) condition number of the overlap matrix $\mathbf{S}$, and (c) variance of the approximate wavefunction as a function of time steps, $n$. The results are shown for the water molecule for the extended bond length, $R=2.8$ a.u., from Figure 1-b. The operator, $f(\hat{H}) = e^{-i\hat{H}\tau}$, was used for the KDM method (blue) and FDM method (red). The FDM results are decomposed in terms of various total number of filter energies ($J=3,4,5$) for the small frequency window and $J=6,7,8$ for the large frequency window. The transparency of the KDM curve is determined by a threshold value, $\kappa_{th} = 1\times 10^{17}$, for illustration purposes, where the dark blue lines correspond to $\kappa <  \kappa_{th}$. }
    \label{fig:Convergence}
\end{figure}

To better compare the KDM and FDM-based algorithms, we plot the energy error, condition number, and variance in Figure 3 as a function of total number of time steps. We only plot the results for the water molecule case from Figure 1-b at the long interatomic distance of $R=2.8$ a.u. because it represents the toughest point with the largest energy error for both methods. In general, we found similar behavior for the other two molecules for different interatomic distances. In the Figure, we only plot the $f(\hat{H}) = e^{-i\hat{H}\tau}$ operator case, though similar behavior is seen for the Hamiltonian case, $f(\hat{H}) = \hat{H}$. In the top row, we present the result for the narrow filter frequency window, $[E_{\text{HF}}- 0.3 E_h,E_{\text{HF}}+ 0.2 E_h]$, while the bottom row presents the results for the large frequency window, $[E_{\text{HF}}- 20 E_h,E_{\text{HF}}+ 20 E_h]$. We emphasize that we have not performed the hyperparameter optimization loop as discussed above, but chose the appropriate frequency window and total number of filter energies $J$ for illustration purposes. We first analyze the KDM U method results which is the same for the top and bottom rows. We find that the KDM U method reaches chemical accuracy within four time steps while also reaching the minimum variance of $2\times 10^{-6}$ within eight time steps. The condition number of the KDM method, however, displays a troubling increasing trend as a function of time steps. We highlight this point by controlling the transparency of the KDM curve based on the threshold value, $\kappa_{th} = 1\times 10^{17}$, where the dark blue lines correspond to $\kappa < \kappa_{th}$. In general, we find that the KDM method becomes unstable as the number of time steps increases, as shown by the fluctuating energy error (light blue lines) and relatively large variance values $\sim 10^{-2}$. Comparatively, the FDM method displays drastically different behavior based on the chosen frequency window. While the small frequency window stabilizes the FDM method quite well, as shown by the decreasing condition numbers in the top row of Figure 3-b, the corresponding energy error (Figure 3-a) and variance (Figure 3-c) decreases much more slowly, never quite reaching the minimum energy error and variance values of the KDM method. It is also worth noting the convergence behavior for the different numbers of filter energies ($J=3,4,5$). The overall condition number for the $J=3$ FDM method is much smaller than the $J=5$ case, though for a large number of time steps, the energy error for the latter is over two orders of magnitude smaller. On the other hand, the FDM method, based on the large frequency window, is able to achieve fast convergence comparable to the KDM method. As the total number of time steps increases, the FDM method displays a much smaller energy error, condition number, and variance compared to the KDM method. These results highlight the fact that for certain problems of interest, where a large number of time steps are required to estimate the ground (or excited-state) energies, the FDM method might be preferable when compared to the KDM method, resulting in much smaller energy errors even within 10 to 20 time steps. 

\begin{figure}
    \centering
    \includegraphics[width=8.9cm]{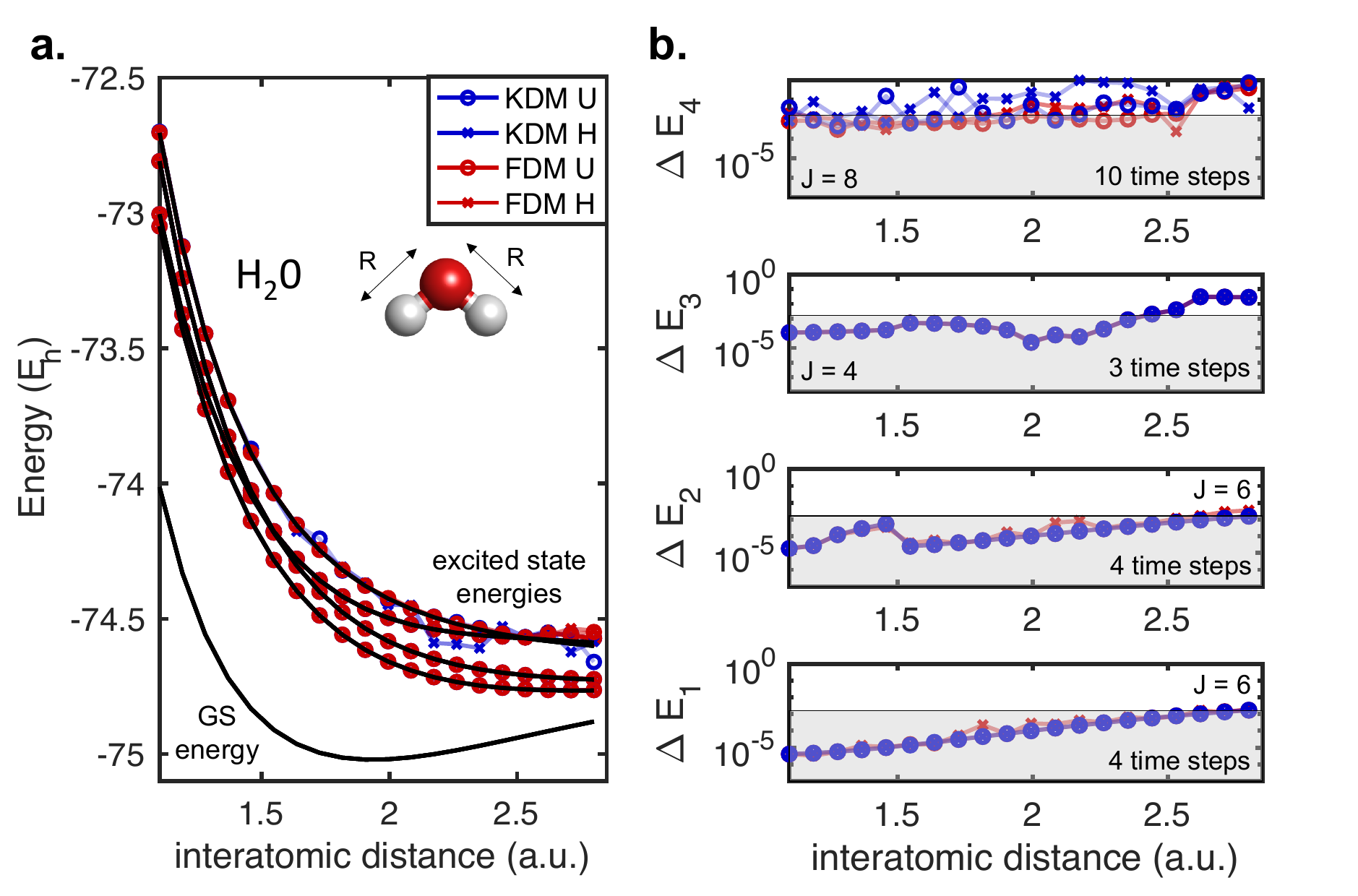}
    \caption{Comparison of KDM and FDM methods for the first four singlet excited state energies of H$_2$O as a function of interatomic distance $R$ between the oxygen and hydrogen atoms. (a) Absolute energy scale (measured in Hartrees) as a function of interatomic distance. (b) Energy error of the $n$th excited state, $\Delta E_n = |E^{(n)}_{\text{FCI}}-E^{(n)}_{\text{approx}}|$, as a function of interatomic distance.}
    \label{fig:ExcitedStates}
\end{figure}

In Figure 4, we numerically simulate the estimation of the first four singlet excited-state energies for the water molecule as a function of interatomic distance $R$. It is important to emphasize that the excited-state energy estimation problem has a wide variety of important applications across physics and chemistry, where classical algorithms are often much less developed than the corresponding ground-state energy estimation algorithms. For illustration purposes, we only consider the singlet ($S=0$) excited-state energy levels, and use singlet states with zero total angular momentum as the initial starting states (see Appendix E for more details) where a time step size of $\tau = 0.1$ a.u. is used for all of the simulations. We emphasize that our  approach does not require knowledge of the ground-state wavefunction (or any other eigenfunction), making this approach distinct from conventional excited-state energy estimation algorithms, specifically within the scope of variational quantum algorithms, which typically use the ground-state wavefunction as a prerequisite. Here, we present the results for the FDM method with a large frequency window. We found that a small frequency window did not give very good results for the FDM method. We find that both the KDM (blue) and FDM (red) methods perform comparably well for the first three excited states. For the first two excited-state energy levels, both methods achieve chemical accuracy within 4 time steps across all interatomic distances. The third and fourth excited state energy curves are quite challenging due to an avoided crossing at approximately $R=2.6$ a.u. which results in nearly degenerate energy curves for the last four data points. We find that both the KDM and FDM methods achieve chemical accuracy for the third excited-state energy curve for nearly all of the interatomic distances, all within three time steps. On the other hand, the fourth excited state energy curve estimation was particularly challenging for both methods. Although it is not shown here, we found that the KDM method only remained stable up to the fifth time step. However, even within five time steps it was unable to reach chemical accuracy for any of the interatomic distances. For this reason, we present the excited-state energy predictions for the tenth time step, where the KDM U method reaches chemical accuracy for nine out of the twenty data points. Furthermore, as evidenced by Figure 4-a and Figure 4-b, the KDM H method, denoted by the blue crosses, displays a large amount instability where it is only able to reach chemical accuracy for four out of the twenty data points. For the fourth excited-state energy curve, the FDM U method outperforms all of the other methods, achieving chemical accuracy for seventeen out of the twenty data points within 10 time steps. The FDM H method is only able to reach chemical accuracy for nine out of the twenty data points within ten time steps.  

Finally, it is worth emphasizing that increasing the total number of time steps does not help improve the convergence of the KDM method due the large condition numbers, however, using step sizes that are much larger (e.g. $\tau=0.5$ a.u.) does improve the performance of the KDM method quite dramatically. This, however, comes at a cost of requiring larger Trotter circuits for a single time step to maintain an equivalent Trotter error. We should also point out that within all of the numerical experiments, the presented results correspond to a fixed number of time steps across all interatomic distances. In practice, the quantum-classical algorithm could have a predetermined stopping criterion, as illustrated in Figure 1, which would imply that every distance point could terminate with a different number of time steps. This would also hold for the hyperparameter optimization loop, which may be implemented independently for every time step and distance point. In summary, while both methods provide adequate excited-state energy estimation, the FDM method might be the preferable choice and perhaps even the only option for reaching a high level of accuracy based on these numerical simulations. 
\vspace{-0.3cm}
\section*{Discussion}

\paragraph{Single Fidelity Estimation Protocols.}
In the following, we discuss various single fidelity estimation protocols that are relevant to the MFE protocol presented above. Over the past few years, there have been a wide range of resource-efficient fidelity estimation protocols that have been developed for the problem of measuring the state fidelity between two pure states $\ket{\phi_i}$ and $\ket{\phi_j}$. Here, we highlight a few of these methods which could be used to provide a substantial reduction in the overall circuit depth in conjunction with the multi-fidelity estimation protocol, specifically when compared to the standard Hadamard test. Before discussing modern approaches, it is worth noting that in the context of fault-tolerant quantum computing, it is sufficient to use the {SWAP} test to estimate the state fidelity $|\!\braket{\phi_i|\phi_j}\!|^2$, to precision $\epsilon$ with $\mathcal{O}(1/\epsilon^2)$ repeated measurements. However, the original implementation of the {SWAP} test requires a $2N+1$ qubit circuit, where two $N$-qubit registers are used for storing $\ket{\phi_i}$ and $\ket{\phi_j}$ with a controlled-SWAP gate that uses an additional ancilla qubit. 

\subsubsection{Destructive SWAP Test}
As shown in \cite{garcia2013swap,cincio2018learning}, the destructive {SWAP} test achieves the same outcome as the original SWAP test but without the use of an ancilla qubit. This approach uses set of parallel Bell measurements and classical post-processing to achieve the same result as the original {SWAP} test with constant depth, $\mathcal{O}(1)$. As a concrete example, we provide the quantum circuit that implements the destructive {SWAP} test for two 3-qubit registers:
\[
\Qcircuit @C=2em @R=1em @!R {
\lstick{\ket{0}} & \multigate{2}{U_j}  & \ctrl{4} & \gate{H} &  \qw & \qw &\meter\\
\lstick{\ket{0}} & \ghost{U_j} & \qw &  \ctrl{4} &  \gate{H} & \qw &\meter \\
\lstick{\ket{0}} & \ghost{U_j} & \qw &  \qw & \ctrl{4} & \gate{H} &\meter \\
\\
\lstick{\ket{0}} & \multigate{2}{U_i} & \targ & \qw & \qw & \qw&\meter \\
\lstick{\ket{0}} & \ghost{U_i} & \qw & \targ  & \qw & \qw&\meter \\
\lstick{\ket{0}} & \ghost{U_i} & \qw & \qw & \targ  & \qw &\meter
}
\]
These Bell measurements only require one-to-one connectivity between each of the qubits in each of the two registers (as shown above). In principle, this might be a more hardware-friendly connectivity compared to the Hadamard test quantum circuit which requires one-to-all connectivity between the ancilla qubit and the rest of the $N$-qubit register. In addition, the destructive {SWAP} test also allows for an additional reduction in the gate depth by allowing the implementation of Trotterized quantum circuits where only $n/2$ time steps are implemented on each register. Overall, the destructive {SWAP} test avoids the use of controlled multi-qubit unitaries, and reduces the depth of the time-stepping circuits by a half, resulting in a substantial reduction in overall circuit depth compared to the original Hadamard test. 

\subsubsection{Mirror-type Quantum Circuits}
As discussed in \cite{havlivcek2019supervised}, it is also possible to estimate pure-state fidelities by using the following mirror-type quantum circuits (shown for a 3-qubit register):
\[
\begin{array}{c}
\Qcircuit @C=1em @R=.8em {
\lstick{\ket{0}} & \multigate{2}{U_j}  &  \multigate{2}{U_i^\dagger} & \qw & \meter \\
\lstick{\ket{0}} & \ghost{U_j} & \ghost{U^\dag} &\qw & \meter\\
\lstick{\ket{0}} &\ghost{U_j} & \ghost{U^\dag} &\qw & \meter
}
\end{array}
\]
By assuming that the all-zero state $\ket{0}^{\otimes N}$ is the initial state of the quantum circuit, it is then possible to estimate $F_1$ by measuring the transition probability of returning to the same state at the output, i.e. measuring $F_1 = |\!\braket{0|U_i^\dagger U_j|0}\!|^2$. The same type of circuit can be used to measure the second fidelity $F_2$, however, $U_i$ and $U_j$ must be replaced with the appropriate circuits that prepare $\tfrac{1}{\sqrt{2}}(\ket{0}^{\otimes N} + \ket{\phi_i})$ and $\tfrac{1}{\sqrt{2}}(\ket{0}^{\otimes N} + \ket{\phi_j})$ respectively. While this approach requires higher depth quantum circuits, it does not require a second qubit register as in the destructive \texttt{SWAP} test. 

\subsubsection{Randomized Measurements}
Fidelity estimation protocols involving randomized measurements \cite{flammia2011direct,elben2019statistical,elben2020cross,zhu2021cross,huang2020predicting} may also be used for the proposed quantum Krylov subspace algorithms. These types of protocols simply require preparing the quantum state $\ket{\phi_k} = U_k\ket{0}^{\otimes N}$ ($k=i,j$), and applying a randomized measurement quantum circuit at the output, labelled $U_{\text{meas}}$, as shown below for a three-qubit register. 
\[
\begin{array}{c}
\Qcircuit @C=1em @R=.8em {
\lstick{\ket{0}} & \multigate{2}{U_k}  &  \multigate{2}{U_{\text{meas}}} & \qw & \meter \\
\lstick{\ket{0}} & \ghost{U_k} & \ghost{U_{\text{meas}}} &\qw & \meter\\
\lstick{\ket{0}} & \ghost{U_k} & \ghost{U_{\text{meas}}} &\qw & \meter
}
\end{array}
\]
The choice of measurement quantum circuit will result in different approximate classical descriptions of the quantum state $\ket{\phi_k}$, a so-called \emph{classical shadow}, which can be used to estimate an exponential number of local observables as well as state fidelities with a small number of measurements \cite{huang2020predicting}. For instance, it was shown in \cite{huang2020predicting} that $N$-qubit Clifford circuits achieve a precision $\epsilon$ with $1/\epsilon^2$ measurements, while a measurement circuit consisting of randomized single-qubit Pauli gates achieves a precision $\epsilon$ with $4^k/\epsilon^2$ measurements, i.e. scaling exponentially with the locality $k$ of the observable which is relevant for fidelity estimation. While the Clifford circuit approach is the most powerful, it comes with increased depth requirements, making it less desirable for near-term quantum computing applications. In principle, this randomized measurement approach can reduce the number of calls/measurements $N_K$ even further from the estimate that we provided in Table 1, though a more careful analysis of the trade-off between the circuit depth (required for Clifford circuits) and exponential number of samples (as required by the Pauli circuits) would need to be performed in order to understand the benefits of this approach in the near term.


\section*{Concluding Remarks}
Building on important previous work on the eigenpair problem on quantum computers \cite{parrish2019quantum,stair2020multireference}, and
the classical filter diagonalization method \cite{wall1995extraction,mandelshtam1997low,mandelshtam2001fdm}, we have presented a unified view of quantum subspace diagonalization methods including the introduction of three new generalized eigenvalue problems that can be used to estimate the ground and excited-state energies of quantum many-body Hamiltonians. Numerical illustrations of the approaches were carried out for three quantum chemistry problems. Each of these methods provide various advantages and disadvantages in terms of number of calls to the quantum computer, gate depth, classical complexity, and numerical stability on the classical computer portion of the algorithms. A key new aspect of our approach is a multi-fidelity estimation protocol that avoids the use of the Hadamard test to estimate the off-diagonal subspace matrix elements, which allows for a substantial reduction in gate depth. 

Overall, we observed that all of the methods, when tuned properly, converged to the correct eigenstate and provided excellent estimates of the energy eigenvalues with a small number time steps for both the ground and excited state energy estimation problems. While the KDM H and FDM H methods require less circuit depth (equivalent to a single Trotter time step), the KDM U and FDM U methods require a lot less calls to the quantum computer, scaling as $\mathcal{O}(M/\epsilon^2)$ compared to the $\mathcal{O}(LM^p/\epsilon^2)$ scaling. This difference is particularly striking for quantum chemistry Hamiltonians in a localized basis where $L$ scales as $\mathcal{O}(N^4)$ where $N$ is equal to the total number of spin-orbitals. For instance, a 50 spin-orbital problem may have approximately $10^6$ terms in the Hamiltonian, which is substantial. This not only increases the total time to solution, but also increases the overall cost of the algorithm when the quantum computer requires a per-shot fee. 

We also found that by properly tuning the FDM hyperparameters, both FDM methods converged as quickly as the KDM methods while also remaining much more stable over a large number of time steps. As a result, we found that the FDM algorithms achieved a much smaller energy error and variance magnitude compared to the KDM methods as the number of time steps increased. The FDM methods also provide a clear advantage in their classical complexity which might make them more appealing for much larger problems. For example, it is quite possible that a multi-reference method with tens of thousands (or over a million) initial reference states could be used to ensure that the initial wavefunction has a strong overlap with the true ground state wavefunction. In such a case, the classical cost of solving the generalized eigenvalue problem would become a severe limitation for the KDM methods, while the FDM methods would not suffer from such problems. The FDM method should also provide an advantage in finding interior excited-state energies (eigenvalues) where a large number of time steps would be required and the filtering technique becomes more important. 

It should be noted that the QKSD algorithms, and in particular the ones
we introduced
based on diagonalization of the time evolution propagator (the ``U'' forms) share
some commonalities with the quantum phase estimation (QPE) algorithm, as also pointed by Klymko \emph{et al.} \cite{klymko2021real}.
QPE is similar in spirit to Fourier analysis of a correlation function and as a consequence one often sees
arguments that the maximum propagation times (equivalent to maximum circuit depth) must be on the order of 
of $\mathcal{O}(1/\epsilon)$ 
to achieve an eigenvalue estimation accurate to
$\epsilon$ \cite{alan_2020_review,garnet_2020_review}.
This would suggest propagation times several orders of magnitude
greater would be required for QPE algorithms compared to the short propagation times shown here for QKSD algorithms, which was also confirmed in \cite{klymko2021real}.
Variations on QPE that use 
more sophisticated signal processing ideas \cite{O_Brien_2019, Somma_2019}
could potentially overcome
the $\mathcal{O}(1/\epsilon)$ limit. Such methods effectively use prior knowledge about the form of the correlation function or related quantities and, as a result, the QKSD approaches discussed here
could also be put in this class. 
 
Lastly, it is worth mentioning that the proposed algorithms can also be used in conjunction with variational quantum algorithms that use parameterized quantum circuits. This approach would use the final optimized quantum circuit, obtained from a specific variational quantum algorithm, as the initial state $\ket{\phi_o}$ and would subsequently build a Krylov basis to improve the energy or cost function estimate. A more detailed analysis of the effect of Trotter error, noise, as well as the development of algorithmic-specific error mitigation strategies might also represent important next steps for future work.

\section{ACKNOWLEDGMENTS}
\noindent This material is based upon work supported by Laboratory Directed Research and Development (LDRD) funding from Argonne National Laboratory, provided by the Director, Office of Science, of the U.S. Department of Energy under Contract No. DE-AC02-06CH11357. 
Use of the Center for Nanoscale Materials, an Office of Science user facility, was supported by the U.S. Department of Energy, Office of Science, Office of Basic Energy Sciences, under Contract No. DE-AC02-06CH11357. We thank Dmitry A. Fedorov and Yuri Alexeev for helpful discussions.

\bibliography{REFERENCES.bib}

\begin{thebibliography}{56}%
\makeatletter
\providecommand \@ifxundefined [1]{%
 \@ifx{#1\undefined}
}%
\providecommand \@ifnum [1]{%
 \ifnum #1\expandafter \@firstoftwo
 \else \expandafter \@secondoftwo
 \fi
}%
\providecommand \@ifx [1]{%
 \ifx #1\expandafter \@firstoftwo
 \else \expandafter \@secondoftwo
 \fi
}%
\providecommand \natexlab [1]{#1}%
\providecommand \enquote  [1]{``#1''}%
\providecommand \bibnamefont  [1]{#1}%
\providecommand \bibfnamefont [1]{#1}%
\providecommand \citenamefont [1]{#1}%
\providecommand \href@noop [0]{\@secondoftwo}%
\providecommand \href [0]{\begingroup \@sanitize@url \@href}%
\providecommand \@href[1]{\@@startlink{#1}\@@href}%
\providecommand \@@href[1]{\endgroup#1\@@endlink}%
\providecommand \@sanitize@url [0]{\catcode `\\12\catcode `\$12\catcode
  `\&12\catcode `\#12\catcode `\^12\catcode `\_12\catcode `\%12\relax}%
\providecommand \@@startlink[1]{}%
\providecommand \@@endlink[0]{}%
\providecommand \url  [0]{\begingroup\@sanitize@url \@url }%
\providecommand \@url [1]{\endgroup\@href {#1}{\urlprefix }}%
\providecommand \urlprefix  [0]{URL }%
\providecommand \Eprint [0]{\href }%
\providecommand \doibase [0]{http://dx.doi.org/}%
\providecommand \selectlanguage [0]{\@gobble}%
\providecommand \bibinfo  [0]{\@secondoftwo}%
\providecommand \bibfield  [0]{\@secondoftwo}%
\providecommand \translation [1]{[#1]}%
\providecommand \BibitemOpen [0]{}%
\providecommand \bibitemStop [0]{}%
\providecommand \bibitemNoStop [0]{.\EOS\space}%
\providecommand \EOS [0]{\spacefactor3000\relax}%
\providecommand \BibitemShut  [1]{\csname bibitem#1\endcsname}%
\let\auto@bib@innerbib\@empty
\bibitem [{\citenamefont {Lloyd}(1996)}]{lloyd1996universal}%
  \BibitemOpen
  \bibfield  {author} {\bibinfo {author} {\bibfnamefont {Seth}\ \bibnamefont
  {Lloyd}},\ }\bibfield  {title} {\enquote {\bibinfo {title} {Universal quantum
  simulators},}\ }\href@noop {} {\bibfield  {journal} {\bibinfo  {journal}
  {Science}\ ,\ \bibinfo {pages} {1073--1078}} (\bibinfo {year}
  {1996})}\BibitemShut {NoStop}%
\bibitem [{\citenamefont {Abrams}\ and\ \citenamefont
  {Lloyd}(1997)}]{abrams1997simulation}%
  \BibitemOpen
  \bibfield  {author} {\bibinfo {author} {\bibfnamefont {Daniel~S}\
  \bibnamefont {Abrams}}\ and\ \bibinfo {author} {\bibfnamefont {Seth}\
  \bibnamefont {Lloyd}},\ }\bibfield  {title} {\enquote {\bibinfo {title}
  {Simulation of many-body fermi systems on a universal quantum computer},}\
  }\href@noop {} {\bibfield  {journal} {\bibinfo  {journal} {Physical Review
  Letters}\ }\textbf {\bibinfo {volume} {79}},\ \bibinfo {pages} {2586}
  (\bibinfo {year} {1997})}\BibitemShut {NoStop}%
\bibitem [{\citenamefont {Zalka}(1998)}]{zalka1998efficient}%
  \BibitemOpen
  \bibfield  {author} {\bibinfo {author} {\bibfnamefont {Christof}\
  \bibnamefont {Zalka}},\ }\bibfield  {title} {\enquote {\bibinfo {title}
  {Efficient simulation of quantum systems by quantum computers},}\ }\href@noop
  {} {\bibfield  {journal} {\bibinfo  {journal} {Fortschritte der Physik:
  Progress of Physics}\ }\textbf {\bibinfo {volume} {46}},\ \bibinfo {pages}
  {877--879} (\bibinfo {year} {1998})}\BibitemShut {NoStop}%
\bibitem [{\citenamefont {Aspuru-Guzik}\ \emph {et~al.}(2005)\citenamefont
  {Aspuru-Guzik}, \citenamefont {Dutoi}, \citenamefont {Love},\ and\
  \citenamefont {Head-Gordon}}]{aspuru2005simulated}%
  \BibitemOpen
  \bibfield  {author} {\bibinfo {author} {\bibfnamefont {Al{\'a}n}\
  \bibnamefont {Aspuru-Guzik}}, \bibinfo {author} {\bibfnamefont {Anthony~D}\
  \bibnamefont {Dutoi}}, \bibinfo {author} {\bibfnamefont {Peter~J}\
  \bibnamefont {Love}}, \ and\ \bibinfo {author} {\bibfnamefont {Martin}\
  \bibnamefont {Head-Gordon}},\ }\bibfield  {title} {\enquote {\bibinfo {title}
  {Simulated quantum computation of molecular energies},}\ }\href@noop {}
  {\bibfield  {journal} {\bibinfo  {journal} {Science}\ }\textbf {\bibinfo
  {volume} {309}},\ \bibinfo {pages} {1704--1707} (\bibinfo {year}
  {2005})}\BibitemShut {NoStop}%
\bibitem [{\citenamefont {Nielsen}\ and\ \citenamefont
  {Chuang}(2002)}]{nielsen2002quantum}%
  \BibitemOpen
  \bibfield  {author} {\bibinfo {author} {\bibfnamefont {Michael~A}\
  \bibnamefont {Nielsen}}\ and\ \bibinfo {author} {\bibfnamefont {Isaac}\
  \bibnamefont {Chuang}},\ }\href@noop {} {\enquote {\bibinfo {title} {Quantum
  computation and quantum information},}\ } (\bibinfo {year}
  {2002})\BibitemShut {NoStop}%
\bibitem [{\citenamefont {Kitaev}(1997)}]{kitaev1997quantum}%
  \BibitemOpen
  \bibfield  {author} {\bibinfo {author} {\bibfnamefont {Aleksei~Yur'evich}\
  \bibnamefont {Kitaev}},\ }\bibfield  {title} {\enquote {\bibinfo {title}
  {Quantum computations: algorithms and error correction},}\ }\href@noop {}
  {\bibfield  {journal} {\bibinfo  {journal} {Uspekhi Matematicheskikh Nauk}\
  }\textbf {\bibinfo {volume} {52}},\ \bibinfo {pages} {53--112} (\bibinfo
  {year} {1997})}\BibitemShut {NoStop}%
\bibitem [{\citenamefont {Cleve}\ \emph {et~al.}(1998)\citenamefont {Cleve},
  \citenamefont {Ekert}, \citenamefont {Macchiavello},\ and\ \citenamefont
  {Mosca}}]{cleve1998quantum}%
  \BibitemOpen
  \bibfield  {author} {\bibinfo {author} {\bibfnamefont {Richard}\ \bibnamefont
  {Cleve}}, \bibinfo {author} {\bibfnamefont {Artur}\ \bibnamefont {Ekert}},
  \bibinfo {author} {\bibfnamefont {Chiara}\ \bibnamefont {Macchiavello}}, \
  and\ \bibinfo {author} {\bibfnamefont {Michele}\ \bibnamefont {Mosca}},\
  }\bibfield  {title} {\enquote {\bibinfo {title} {Quantum algorithms
  revisited},}\ }\href@noop {} {\bibfield  {journal} {\bibinfo  {journal}
  {Proceedings of the Royal Society of London. Series A: Mathematical, Physical
  and Engineering Sciences}\ }\textbf {\bibinfo {volume} {454}},\ \bibinfo
  {pages} {339--354} (\bibinfo {year} {1998})}\BibitemShut {NoStop}%
\bibitem [{\citenamefont {Peruzzo}\ \emph {et~al.}(2014)\citenamefont
  {Peruzzo}, \citenamefont {McClean}, \citenamefont {Shadbolt}, \citenamefont
  {Yung}, \citenamefont {Zhou}, \citenamefont {Love}, \citenamefont
  {Aspuru-Guzik},\ and\ \citenamefont {O’brien}}]{peruzzo2014variational}%
  \BibitemOpen
  \bibfield  {author} {\bibinfo {author} {\bibfnamefont {Alberto}\ \bibnamefont
  {Peruzzo}}, \bibinfo {author} {\bibfnamefont {Jarrod}\ \bibnamefont
  {McClean}}, \bibinfo {author} {\bibfnamefont {Peter}\ \bibnamefont
  {Shadbolt}}, \bibinfo {author} {\bibfnamefont {Man-Hong}\ \bibnamefont
  {Yung}}, \bibinfo {author} {\bibfnamefont {Xiao-Qi}\ \bibnamefont {Zhou}},
  \bibinfo {author} {\bibfnamefont {Peter~J}\ \bibnamefont {Love}}, \bibinfo
  {author} {\bibfnamefont {Al{\'a}n}\ \bibnamefont {Aspuru-Guzik}}, \ and\
  \bibinfo {author} {\bibfnamefont {Jeremy~L}\ \bibnamefont {O’brien}},\
  }\bibfield  {title} {\enquote {\bibinfo {title} {A variational eigenvalue
  solver on a photonic quantum processor},}\ }\href@noop {} {\bibfield
  {journal} {\bibinfo  {journal} {Nature Communications}\ }\textbf {\bibinfo
  {volume} {5}},\ \bibinfo {pages} {1--7} (\bibinfo {year} {2014})}\BibitemShut
  {NoStop}%
\bibitem [{\citenamefont {McClean}\ \emph {et~al.}(2016)\citenamefont
  {McClean}, \citenamefont {Romero}, \citenamefont {Babbush},\ and\
  \citenamefont {Aspuru-Guzik}}]{mcclean2016theory}%
  \BibitemOpen
  \bibfield  {author} {\bibinfo {author} {\bibfnamefont {Jarrod~R}\
  \bibnamefont {McClean}}, \bibinfo {author} {\bibfnamefont {Jonathan}\
  \bibnamefont {Romero}}, \bibinfo {author} {\bibfnamefont {Ryan}\ \bibnamefont
  {Babbush}}, \ and\ \bibinfo {author} {\bibfnamefont {Al{\'a}n}\ \bibnamefont
  {Aspuru-Guzik}},\ }\bibfield  {title} {\enquote {\bibinfo {title} {The theory
  of variational hybrid quantum-classical algorithms},}\ }\href@noop {}
  {\bibfield  {journal} {\bibinfo  {journal} {New Journal of Physics}\ }\textbf
  {\bibinfo {volume} {18}},\ \bibinfo {pages} {023023} (\bibinfo {year}
  {2016})}\BibitemShut {NoStop}%
\bibitem [{\citenamefont {Kandala}\ \emph {et~al.}(2017)\citenamefont
  {Kandala}, \citenamefont {Mezzacapo}, \citenamefont {Temme}, \citenamefont
  {Takita}, \citenamefont {Brink}, \citenamefont {Chow},\ and\ \citenamefont
  {Gambetta}}]{kandala2017hardware}%
  \BibitemOpen
  \bibfield  {author} {\bibinfo {author} {\bibfnamefont {Abhinav}\ \bibnamefont
  {Kandala}}, \bibinfo {author} {\bibfnamefont {Antonio}\ \bibnamefont
  {Mezzacapo}}, \bibinfo {author} {\bibfnamefont {Kristan}\ \bibnamefont
  {Temme}}, \bibinfo {author} {\bibfnamefont {Maika}\ \bibnamefont {Takita}},
  \bibinfo {author} {\bibfnamefont {Markus}\ \bibnamefont {Brink}}, \bibinfo
  {author} {\bibfnamefont {Jerry~M}\ \bibnamefont {Chow}}, \ and\ \bibinfo
  {author} {\bibfnamefont {Jay~M}\ \bibnamefont {Gambetta}},\ }\bibfield
  {title} {\enquote {\bibinfo {title} {Hardware-efficient variational quantum
  eigensolver for small molecules and quantum magnets},}\ }\href@noop {}
  {\bibfield  {journal} {\bibinfo  {journal} {Nature}\ }\textbf {\bibinfo
  {volume} {549}},\ \bibinfo {pages} {242--246} (\bibinfo {year}
  {2017})}\BibitemShut {NoStop}%
\bibitem [{\citenamefont {Farhi}\ \emph {et~al.}(2014)\citenamefont {Farhi},
  \citenamefont {Goldstone},\ and\ \citenamefont {Gutmann}}]{farhi2014quantum}%
  \BibitemOpen
  \bibfield  {author} {\bibinfo {author} {\bibfnamefont {Edward}\ \bibnamefont
  {Farhi}}, \bibinfo {author} {\bibfnamefont {Jeffrey}\ \bibnamefont
  {Goldstone}}, \ and\ \bibinfo {author} {\bibfnamefont {Sam}\ \bibnamefont
  {Gutmann}},\ }\bibfield  {title} {\enquote {\bibinfo {title} {A quantum
  approximate optimization algorithm},}\ }\href@noop {} {\bibfield  {journal}
  {\bibinfo  {journal} {arXiv preprint arXiv:1411.4028}\ } (\bibinfo {year}
  {2014})}\BibitemShut {NoStop}%
\bibitem [{\citenamefont {Preskill}(2018)}]{preskill2018quantum}%
  \BibitemOpen
  \bibfield  {author} {\bibinfo {author} {\bibfnamefont {John}\ \bibnamefont
  {Preskill}},\ }\bibfield  {title} {\enquote {\bibinfo {title} {Quantum
  computing in the nisq era and beyond},}\ }\href@noop {} {\bibfield  {journal}
  {\bibinfo  {journal} {Quantum}\ }\textbf {\bibinfo {volume} {2}},\ \bibinfo
  {pages} {79} (\bibinfo {year} {2018})}\BibitemShut {NoStop}%
\bibitem [{\citenamefont {Cerezo}\ \emph
  {et~al.}(2021{\natexlab{a}})\citenamefont {Cerezo}, \citenamefont
  {Arrasmith}, \citenamefont {Babbush}, \citenamefont {Benjamin}, \citenamefont
  {Endo}, \citenamefont {Fujii}, \citenamefont {McClean}, \citenamefont
  {Mitarai}, \citenamefont {Yuan}, \citenamefont {Cincio} \emph
  {et~al.}}]{cerezo2021variational}%
  \BibitemOpen
  \bibfield  {author} {\bibinfo {author} {\bibfnamefont {Marco}\ \bibnamefont
  {Cerezo}}, \bibinfo {author} {\bibfnamefont {Andrew}\ \bibnamefont
  {Arrasmith}}, \bibinfo {author} {\bibfnamefont {Ryan}\ \bibnamefont
  {Babbush}}, \bibinfo {author} {\bibfnamefont {Simon~C}\ \bibnamefont
  {Benjamin}}, \bibinfo {author} {\bibfnamefont {Suguru}\ \bibnamefont {Endo}},
  \bibinfo {author} {\bibfnamefont {Keisuke}\ \bibnamefont {Fujii}}, \bibinfo
  {author} {\bibfnamefont {Jarrod~R}\ \bibnamefont {McClean}}, \bibinfo
  {author} {\bibfnamefont {Kosuke}\ \bibnamefont {Mitarai}}, \bibinfo {author}
  {\bibfnamefont {Xiao}\ \bibnamefont {Yuan}}, \bibinfo {author} {\bibfnamefont
  {Lukasz}\ \bibnamefont {Cincio}},  \emph {et~al.},\ }\bibfield  {title}
  {\enquote {\bibinfo {title} {Variational quantum algorithms},}\ }\href@noop
  {} {\bibfield  {journal} {\bibinfo  {journal} {Nature Reviews Physics}\ ,\
  \bibinfo {pages} {1--20}} (\bibinfo {year} {2021}{\natexlab{a}})}\BibitemShut
  {NoStop}%
\bibitem [{\citenamefont {Santagati}\ \emph {et~al.}(2018)\citenamefont
  {Santagati}, \citenamefont {Wang}, \citenamefont {Gentile}, \citenamefont
  {Paesani}, \citenamefont {Wiebe}, \citenamefont {McClean}, \citenamefont
  {Morley-Short}, \citenamefont {Shadbolt}, \citenamefont {Bonneau},
  \citenamefont {Silverstone} \emph {et~al.}}]{santagati2018witnessing}%
  \BibitemOpen
  \bibfield  {author} {\bibinfo {author} {\bibfnamefont {Raffaele}\
  \bibnamefont {Santagati}}, \bibinfo {author} {\bibfnamefont {Jianwei}\
  \bibnamefont {Wang}}, \bibinfo {author} {\bibfnamefont {Antonio~A}\
  \bibnamefont {Gentile}}, \bibinfo {author} {\bibfnamefont {Stefano}\
  \bibnamefont {Paesani}}, \bibinfo {author} {\bibfnamefont {Nathan}\
  \bibnamefont {Wiebe}}, \bibinfo {author} {\bibfnamefont {Jarrod~R}\
  \bibnamefont {McClean}}, \bibinfo {author} {\bibfnamefont {Sam}\ \bibnamefont
  {Morley-Short}}, \bibinfo {author} {\bibfnamefont {Peter~J}\ \bibnamefont
  {Shadbolt}}, \bibinfo {author} {\bibfnamefont {Damien}\ \bibnamefont
  {Bonneau}}, \bibinfo {author} {\bibfnamefont {Joshua~W}\ \bibnamefont
  {Silverstone}},  \emph {et~al.},\ }\bibfield  {title} {\enquote {\bibinfo
  {title} {Witnessing eigenstates for quantum simulation of hamiltonian
  spectra},}\ }\href@noop {} {\bibfield  {journal} {\bibinfo  {journal}
  {Science advances}\ }\textbf {\bibinfo {volume} {4}},\ \bibinfo {pages}
  {eaap9646} (\bibinfo {year} {2018})}\BibitemShut {NoStop}%
\bibitem [{\citenamefont {Higgott}\ \emph {et~al.}(2019)\citenamefont
  {Higgott}, \citenamefont {Wang},\ and\ \citenamefont
  {Brierley}}]{higgott2019variational}%
  \BibitemOpen
  \bibfield  {author} {\bibinfo {author} {\bibfnamefont {Oscar}\ \bibnamefont
  {Higgott}}, \bibinfo {author} {\bibfnamefont {Daochen}\ \bibnamefont {Wang}},
  \ and\ \bibinfo {author} {\bibfnamefont {Stephen}\ \bibnamefont {Brierley}},\
  }\bibfield  {title} {\enquote {\bibinfo {title} {Variational quantum
  computation of excited states},}\ }\href@noop {} {\bibfield  {journal}
  {\bibinfo  {journal} {Quantum}\ }\textbf {\bibinfo {volume} {3}},\ \bibinfo
  {pages} {156} (\bibinfo {year} {2019})}\BibitemShut {NoStop}%
\bibitem [{\citenamefont {Sim}\ \emph {et~al.}(2018)\citenamefont {Sim},
  \citenamefont {Romero}, \citenamefont {Johnson},\ and\ \citenamefont
  {Aspuru-Guzik}}]{sim2018quantum}%
  \BibitemOpen
  \bibfield  {author} {\bibinfo {author} {\bibfnamefont {Sukin}\ \bibnamefont
  {Sim}}, \bibinfo {author} {\bibfnamefont {Jonathan}\ \bibnamefont {Romero}},
  \bibinfo {author} {\bibfnamefont {Peter~D}\ \bibnamefont {Johnson}}, \ and\
  \bibinfo {author} {\bibfnamefont {Al{\'a}n}\ \bibnamefont {Aspuru-Guzik}},\
  }\bibfield  {title} {\enquote {\bibinfo {title} {Quantum computer simulates
  excited states of molecule},}\ }\href@noop {} {\bibfield  {journal} {\bibinfo
   {journal} {Physics}\ }\textbf {\bibinfo {volume} {11}},\ \bibinfo {pages}
  {14} (\bibinfo {year} {2018})}\BibitemShut {NoStop}%
\bibitem [{\citenamefont {Bittel}\ and\ \citenamefont
  {Kliesch}(2021)}]{bittel2021training}%
  \BibitemOpen
  \bibfield  {author} {\bibinfo {author} {\bibfnamefont {Lennart}\ \bibnamefont
  {Bittel}}\ and\ \bibinfo {author} {\bibfnamefont {Martin}\ \bibnamefont
  {Kliesch}},\ }\bibfield  {title} {\enquote {\bibinfo {title} {Training
  variational quantum algorithms is np-hard--even for logarithmically many
  qubits and free fermionic systems},}\ }\href@noop {} {\bibfield  {journal}
  {\bibinfo  {journal} {arXiv preprint arXiv:2101.07267}\ } (\bibinfo {year}
  {2021})}\BibitemShut {NoStop}%
\bibitem [{\citenamefont {McClean}\ \emph {et~al.}(2018)\citenamefont
  {McClean}, \citenamefont {Boixo}, \citenamefont {Smelyanskiy}, \citenamefont
  {Babbush},\ and\ \citenamefont {Neven}}]{mcclean2018barren}%
  \BibitemOpen
  \bibfield  {author} {\bibinfo {author} {\bibfnamefont {Jarrod~R}\
  \bibnamefont {McClean}}, \bibinfo {author} {\bibfnamefont {Sergio}\
  \bibnamefont {Boixo}}, \bibinfo {author} {\bibfnamefont {Vadim~N}\
  \bibnamefont {Smelyanskiy}}, \bibinfo {author} {\bibfnamefont {Ryan}\
  \bibnamefont {Babbush}}, \ and\ \bibinfo {author} {\bibfnamefont {Hartmut}\
  \bibnamefont {Neven}},\ }\bibfield  {title} {\enquote {\bibinfo {title}
  {Barren plateaus in quantum neural network training landscapes},}\
  }\href@noop {} {\bibfield  {journal} {\bibinfo  {journal} {Nature
  Communications}\ }\textbf {\bibinfo {volume} {9}},\ \bibinfo {pages} {1--6}
  (\bibinfo {year} {2018})}\BibitemShut {NoStop}%
\bibitem [{\citenamefont {Wang}\ \emph {et~al.}(2020)\citenamefont {Wang},
  \citenamefont {Fontana}, \citenamefont {Cerezo}, \citenamefont {Sharma},
  \citenamefont {Sone}, \citenamefont {Cincio},\ and\ \citenamefont
  {Coles}}]{wang2020noise}%
  \BibitemOpen
  \bibfield  {author} {\bibinfo {author} {\bibfnamefont {Samson}\ \bibnamefont
  {Wang}}, \bibinfo {author} {\bibfnamefont {Enrico}\ \bibnamefont {Fontana}},
  \bibinfo {author} {\bibfnamefont {Marco}\ \bibnamefont {Cerezo}}, \bibinfo
  {author} {\bibfnamefont {Kunal}\ \bibnamefont {Sharma}}, \bibinfo {author}
  {\bibfnamefont {Akira}\ \bibnamefont {Sone}}, \bibinfo {author}
  {\bibfnamefont {Lukasz}\ \bibnamefont {Cincio}}, \ and\ \bibinfo {author}
  {\bibfnamefont {Patrick~J}\ \bibnamefont {Coles}},\ }\bibfield  {title}
  {\enquote {\bibinfo {title} {Noise-induced barren plateaus in variational
  quantum algorithms},}\ }\href@noop {} {\bibfield  {journal} {\bibinfo
  {journal} {arXiv preprint arXiv:2007.14384}\ } (\bibinfo {year}
  {2020})}\BibitemShut {NoStop}%
\bibitem [{\citenamefont {Cerezo}\ \emph
  {et~al.}(2021{\natexlab{b}})\citenamefont {Cerezo}, \citenamefont {Sone},
  \citenamefont {Volkoff}, \citenamefont {Cincio},\ and\ \citenamefont
  {Coles}}]{cerezo2021cost}%
  \BibitemOpen
  \bibfield  {author} {\bibinfo {author} {\bibfnamefont {Marco}\ \bibnamefont
  {Cerezo}}, \bibinfo {author} {\bibfnamefont {Akira}\ \bibnamefont {Sone}},
  \bibinfo {author} {\bibfnamefont {Tyler}\ \bibnamefont {Volkoff}}, \bibinfo
  {author} {\bibfnamefont {Lukasz}\ \bibnamefont {Cincio}}, \ and\ \bibinfo
  {author} {\bibfnamefont {Patrick~J}\ \bibnamefont {Coles}},\ }\bibfield
  {title} {\enquote {\bibinfo {title} {Cost function dependent barren plateaus
  in shallow parametrized quantum circuits},}\ }\href@noop {} {\bibfield
  {journal} {\bibinfo  {journal} {Nature Communications}\ }\textbf {\bibinfo
  {volume} {12}},\ \bibinfo {pages} {1--12} (\bibinfo {year}
  {2021}{\natexlab{b}})}\BibitemShut {NoStop}%
\bibitem [{\citenamefont {Arrasmith}\ \emph {et~al.}(2020)\citenamefont
  {Arrasmith}, \citenamefont {Cerezo}, \citenamefont {Czarnik}, \citenamefont
  {Cincio},\ and\ \citenamefont {Coles}}]{arrasmith2020effect}%
  \BibitemOpen
  \bibfield  {author} {\bibinfo {author} {\bibfnamefont {Andrew}\ \bibnamefont
  {Arrasmith}}, \bibinfo {author} {\bibfnamefont {M}~\bibnamefont {Cerezo}},
  \bibinfo {author} {\bibfnamefont {Piotr}\ \bibnamefont {Czarnik}}, \bibinfo
  {author} {\bibfnamefont {Lukasz}\ \bibnamefont {Cincio}}, \ and\ \bibinfo
  {author} {\bibfnamefont {Patrick~J}\ \bibnamefont {Coles}},\ }\bibfield
  {title} {\enquote {\bibinfo {title} {Effect of barren plateaus on
  gradient-free optimization},}\ }\href@noop {} {\bibfield  {journal} {\bibinfo
   {journal} {arXiv preprint arXiv:2011.12245}\ } (\bibinfo {year}
  {2020})}\BibitemShut {NoStop}%
\bibitem [{\citenamefont {Huggins}\ \emph {et~al.}(2020)\citenamefont
  {Huggins}, \citenamefont {Lee}, \citenamefont {Baek}, \citenamefont
  {O’Gorman},\ and\ \citenamefont {Whaley}}]{huggins2020non}%
  \BibitemOpen
  \bibfield  {author} {\bibinfo {author} {\bibfnamefont {William~J}\
  \bibnamefont {Huggins}}, \bibinfo {author} {\bibfnamefont {Joonho}\
  \bibnamefont {Lee}}, \bibinfo {author} {\bibfnamefont {Unpil}\ \bibnamefont
  {Baek}}, \bibinfo {author} {\bibfnamefont {Bryan}\ \bibnamefont
  {O’Gorman}}, \ and\ \bibinfo {author} {\bibfnamefont {K~Birgitta}\
  \bibnamefont {Whaley}},\ }\bibfield  {title} {\enquote {\bibinfo {title} {A
  non-orthogonal variational quantum eigensolver},}\ }\href@noop {} {\bibfield
  {journal} {\bibinfo  {journal} {New Journal of Physics}\ }\textbf {\bibinfo
  {volume} {22}},\ \bibinfo {pages} {073009} (\bibinfo {year}
  {2020})}\BibitemShut {NoStop}%
\bibitem [{\citenamefont {Parrish}\ and\ \citenamefont
  {McMahon}(2019)}]{parrish2019quantum}%
  \BibitemOpen
  \bibfield  {author} {\bibinfo {author} {\bibfnamefont {Robert~M}\
  \bibnamefont {Parrish}}\ and\ \bibinfo {author} {\bibfnamefont {Peter~L}\
  \bibnamefont {McMahon}},\ }\bibfield  {title} {\enquote {\bibinfo {title}
  {Quantum filter diagonalization: Quantum eigendecomposition without full
  quantum phase estimation},}\ }\href@noop {} {\bibfield  {journal} {\bibinfo
  {journal} {arXiv preprint arXiv:1909.08925}\ } (\bibinfo {year}
  {2019})}\BibitemShut {NoStop}%
\bibitem [{\citenamefont {Seki}\ and\ \citenamefont
  {Yunoki}(2021)}]{seki2021quantum}%
  \BibitemOpen
  \bibfield  {author} {\bibinfo {author} {\bibfnamefont {Kazuhiro}\
  \bibnamefont {Seki}}\ and\ \bibinfo {author} {\bibfnamefont {Seiji}\
  \bibnamefont {Yunoki}},\ }\bibfield  {title} {\enquote {\bibinfo {title}
  {Quantum power method by a superposition of time-evolved states},}\
  }\href@noop {} {\bibfield  {journal} {\bibinfo  {journal} {PRX Quantum}\
  }\textbf {\bibinfo {volume} {2}},\ \bibinfo {pages} {010333} (\bibinfo {year}
  {2021})}\BibitemShut {NoStop}%
\bibitem [{\citenamefont {Stair}\ \emph {et~al.}(2020)\citenamefont {Stair},
  \citenamefont {Huang},\ and\ \citenamefont
  {Evangelista}}]{stair2020multireference}%
  \BibitemOpen
  \bibfield  {author} {\bibinfo {author} {\bibfnamefont {Nicholas~H}\
  \bibnamefont {Stair}}, \bibinfo {author} {\bibfnamefont {Renke}\ \bibnamefont
  {Huang}}, \ and\ \bibinfo {author} {\bibfnamefont {Francesco~A}\ \bibnamefont
  {Evangelista}},\ }\bibfield  {title} {\enquote {\bibinfo {title} {A
  multireference quantum krylov algorithm for strongly correlated electrons},}\
  }\href@noop {} {\bibfield  {journal} {\bibinfo  {journal} {Journal of
  Chemical Theory and Computation}\ }\textbf {\bibinfo {volume} {16}},\
  \bibinfo {pages} {2236--2245} (\bibinfo {year} {2020})}\BibitemShut {NoStop}%
\bibitem [{\citenamefont {McClean}\ \emph {et~al.}(2017)\citenamefont
  {McClean}, \citenamefont {Kimchi-Schwartz}, \citenamefont {Carter},\ and\
  \citenamefont {De~Jong}}]{mcclean2017hybrid}%
  \BibitemOpen
  \bibfield  {author} {\bibinfo {author} {\bibfnamefont {Jarrod~R}\
  \bibnamefont {McClean}}, \bibinfo {author} {\bibfnamefont {Mollie~E}\
  \bibnamefont {Kimchi-Schwartz}}, \bibinfo {author} {\bibfnamefont {Jonathan}\
  \bibnamefont {Carter}}, \ and\ \bibinfo {author} {\bibfnamefont {Wibe~A}\
  \bibnamefont {De~Jong}},\ }\bibfield  {title} {\enquote {\bibinfo {title}
  {Hybrid quantum-classical hierarchy for mitigation of decoherence and
  determination of excited states},}\ }\href@noop {} {\bibfield  {journal}
  {\bibinfo  {journal} {Physical Review A}\ }\textbf {\bibinfo {volume} {95}},\
  \bibinfo {pages} {042308} (\bibinfo {year} {2017})}\BibitemShut {NoStop}%
\bibitem [{\citenamefont {Colless}\ \emph {et~al.}(2018)\citenamefont
  {Colless}, \citenamefont {Ramasesh}, \citenamefont {Dahlen}, \citenamefont
  {Blok}, \citenamefont {Kimchi-Schwartz}, \citenamefont {McClean},
  \citenamefont {Carter}, \citenamefont {de~Jong},\ and\ \citenamefont
  {Siddiqi}}]{colless2018computation}%
  \BibitemOpen
  \bibfield  {author} {\bibinfo {author} {\bibfnamefont {James~I}\ \bibnamefont
  {Colless}}, \bibinfo {author} {\bibfnamefont {Vinay~V}\ \bibnamefont
  {Ramasesh}}, \bibinfo {author} {\bibfnamefont {Dar}\ \bibnamefont {Dahlen}},
  \bibinfo {author} {\bibfnamefont {Machiel~S}\ \bibnamefont {Blok}}, \bibinfo
  {author} {\bibfnamefont {Mollie~E}\ \bibnamefont {Kimchi-Schwartz}}, \bibinfo
  {author} {\bibfnamefont {Jarrod~R}\ \bibnamefont {McClean}}, \bibinfo
  {author} {\bibfnamefont {Jonathan}\ \bibnamefont {Carter}}, \bibinfo {author}
  {\bibfnamefont {Wibe~A}\ \bibnamefont {de~Jong}}, \ and\ \bibinfo {author}
  {\bibfnamefont {Irfan}\ \bibnamefont {Siddiqi}},\ }\bibfield  {title}
  {\enquote {\bibinfo {title} {Computation of molecular spectra on a quantum
  processor with an error-resilient algorithm},}\ }\href@noop {} {\bibfield
  {journal} {\bibinfo  {journal} {Physical Review X}\ }\textbf {\bibinfo
  {volume} {8}},\ \bibinfo {pages} {011021} (\bibinfo {year}
  {2018})}\BibitemShut {NoStop}%
\bibitem [{\citenamefont {Bharti}\ and\ \citenamefont
  {Haug}(2020{\natexlab{a}})}]{bharti2020iterative}%
  \BibitemOpen
  \bibfield  {author} {\bibinfo {author} {\bibfnamefont {Kishor}\ \bibnamefont
  {Bharti}}\ and\ \bibinfo {author} {\bibfnamefont {Tobias}\ \bibnamefont
  {Haug}},\ }\bibfield  {title} {\enquote {\bibinfo {title} {Iterative quantum
  assisted eigensolver},}\ }\href@noop {} {\bibfield  {journal} {\bibinfo
  {journal} {arXiv preprint arXiv:2010.05638}\ } (\bibinfo {year}
  {2020}{\natexlab{a}})}\BibitemShut {NoStop}%
\bibitem [{\citenamefont {Bharti}\ and\ \citenamefont
  {Haug}(2020{\natexlab{b}})}]{bharti2020quantum}%
  \BibitemOpen
  \bibfield  {author} {\bibinfo {author} {\bibfnamefont {Kishor}\ \bibnamefont
  {Bharti}}\ and\ \bibinfo {author} {\bibfnamefont {Tobias}\ \bibnamefont
  {Haug}},\ }\bibfield  {title} {\enquote {\bibinfo {title} {Quantum assisted
  simulator},}\ }\href@noop {} {\bibfield  {journal} {\bibinfo  {journal}
  {arXiv preprint arXiv:2011.06911}\ } (\bibinfo {year}
  {2020}{\natexlab{b}})}\BibitemShut {NoStop}%
\bibitem [{\citenamefont {Golub}\ and\ \citenamefont
  {Van~Loan}(1996)}]{GoluVanl96}%
  \BibitemOpen
  \bibfield  {author} {\bibinfo {author} {\bibfnamefont {Gene~H.}\ \bibnamefont
  {Golub}}\ and\ \bibinfo {author} {\bibfnamefont {Charles~F.}\ \bibnamefont
  {Van~Loan}},\ }\href@noop {} {\emph {\bibinfo {title} {Matrix
  Computations}}},\ \bibinfo {edition} {3rd}\ ed.\ (\bibinfo  {publisher} {The
  Johns Hopkins University Press},\ \bibinfo {year} {1996})\BibitemShut
  {NoStop}%
\bibitem [{\citenamefont {Parlett}(1980)}]{Parlett1980}%
  \BibitemOpen
  \bibfield  {author} {\bibinfo {author} {\bibfnamefont {Beresford~N.}\
  \bibnamefont {Parlett}},\ }\href@noop {} {\emph {\bibinfo {title} {The
  Symmetric Eigenvalue Problem}}}\ (\bibinfo  {publisher} {Prentice-Hall},\
  \bibinfo {address} {Englewood Cliffs, N.J.},\ \bibinfo {year}
  {1980})\BibitemShut {NoStop}%
\bibitem [{\citenamefont {Bespalova}\ and\ \citenamefont
  {Kyriienko}(2021)}]{bespalova2021hamiltonian}%
  \BibitemOpen
  \bibfield  {author} {\bibinfo {author} {\bibfnamefont {Tatiana~A}\
  \bibnamefont {Bespalova}}\ and\ \bibinfo {author} {\bibfnamefont {Oleksandr}\
  \bibnamefont {Kyriienko}},\ }\bibfield  {title} {\enquote {\bibinfo {title}
  {Hamiltonian operator approximation for energy measurement and ground-state
  preparation},}\ }\href@noop {} {\bibfield  {journal} {\bibinfo  {journal}
  {PRX Quantum}\ }\textbf {\bibinfo {volume} {2}},\ \bibinfo {pages} {030318}
  (\bibinfo {year} {2021})}\BibitemShut {NoStop}%
\bibitem [{\citenamefont {Motta}\ \emph {et~al.}(2020)\citenamefont {Motta},
  \citenamefont {Sun}, \citenamefont {Tan}, \citenamefont {O’Rourke},
  \citenamefont {Ye}, \citenamefont {Minnich}, \citenamefont {Brand{\~a}o},\
  and\ \citenamefont {Chan}}]{motta2020determining}%
  \BibitemOpen
  \bibfield  {author} {\bibinfo {author} {\bibfnamefont {Mario}\ \bibnamefont
  {Motta}}, \bibinfo {author} {\bibfnamefont {Chong}\ \bibnamefont {Sun}},
  \bibinfo {author} {\bibfnamefont {Adrian~TK}\ \bibnamefont {Tan}}, \bibinfo
  {author} {\bibfnamefont {Matthew~J}\ \bibnamefont {O’Rourke}}, \bibinfo
  {author} {\bibfnamefont {Erika}\ \bibnamefont {Ye}}, \bibinfo {author}
  {\bibfnamefont {Austin~J}\ \bibnamefont {Minnich}}, \bibinfo {author}
  {\bibfnamefont {Fernando~GSL}\ \bibnamefont {Brand{\~a}o}}, \ and\ \bibinfo
  {author} {\bibfnamefont {Garnet Kin-Lic}\ \bibnamefont {Chan}},\ }\bibfield
  {title} {\enquote {\bibinfo {title} {Determining eigenstates and thermal
  states on a quantum computer using quantum imaginary time evolution},}\
  }\href@noop {} {\bibfield  {journal} {\bibinfo  {journal} {Nature Physics}\
  }\textbf {\bibinfo {volume} {16}},\ \bibinfo {pages} {205--210} (\bibinfo
  {year} {2020})}\BibitemShut {NoStop}%
\bibitem [{\citenamefont {Wall}\ and\ \citenamefont
  {Neuhauser}(1995)}]{wall1995extraction}%
  \BibitemOpen
  \bibfield  {author} {\bibinfo {author} {\bibfnamefont {Michael~R}\
  \bibnamefont {Wall}}\ and\ \bibinfo {author} {\bibfnamefont {Daniel}\
  \bibnamefont {Neuhauser}},\ }\bibfield  {title} {\enquote {\bibinfo {title}
  {Extraction, through filter-diagonalization, of general quantum eigenvalues
  or classical normal mode frequencies from a small number of residues or a
  short-time segment of a signal. i. theory and application to a
  quantum-dynamics model},}\ }\href@noop {} {\bibfield  {journal} {\bibinfo
  {journal} {The Journal of Chemical Physics}\ }\textbf {\bibinfo {volume}
  {102}},\ \bibinfo {pages} {8011--8022} (\bibinfo {year} {1995})}\BibitemShut
  {NoStop}%
\bibitem [{\citenamefont {Mandelshtam}\ and\ \citenamefont
  {Taylor}(1997)}]{mandelshtam1997low}%
  \BibitemOpen
  \bibfield  {author} {\bibinfo {author} {\bibfnamefont {Vladimir~A}\
  \bibnamefont {Mandelshtam}}\ and\ \bibinfo {author} {\bibfnamefont
  {Howard~S}\ \bibnamefont {Taylor}},\ }\bibfield  {title} {\enquote {\bibinfo
  {title} {A low-storage filter diagonalization method for quantum eigenenergy
  calculation or for spectral analysis of time signals},}\ }\href@noop {}
  {\bibfield  {journal} {\bibinfo  {journal} {The Journal of Chemical Physics}\
  }\textbf {\bibinfo {volume} {106}},\ \bibinfo {pages} {5085--5090} (\bibinfo
  {year} {1997})}\BibitemShut {NoStop}%
\bibitem [{\citenamefont {Mandelshtam}(2001)}]{mandelshtam2001fdm}%
  \BibitemOpen
  \bibfield  {author} {\bibinfo {author} {\bibfnamefont {Vladimir~A}\
  \bibnamefont {Mandelshtam}},\ }\bibfield  {title} {\enquote {\bibinfo {title}
  {{FDM}: the filter diagonalization method for data processing in nmr
  experiments},}\ }\href@noop {} {\bibfield  {journal} {\bibinfo  {journal}
  {Progress in Nuclear Magnetic Resonance Spectroscopy}\ }\textbf {\bibinfo
  {volume} {38}},\ \bibinfo {pages} {159--196} (\bibinfo {year}
  {2001})}\BibitemShut {NoStop}%
\bibitem [{\citenamefont {Klymko}\ \emph {et~al.}(2021)\citenamefont {Klymko},
  \citenamefont {Mejuto-Zaera}, \citenamefont {Cotton}, \citenamefont
  {Wudarski}, \citenamefont {Urbanek}, \citenamefont {Hait}, \citenamefont
  {Head-Gordon}, \citenamefont {Whaley}, \citenamefont {Moussa}, \citenamefont
  {Wiebe} \emph {et~al.}}]{klymko2021real}%
  \BibitemOpen
  \bibfield  {author} {\bibinfo {author} {\bibfnamefont {Katherine}\
  \bibnamefont {Klymko}}, \bibinfo {author} {\bibfnamefont {Carlos}\
  \bibnamefont {Mejuto-Zaera}}, \bibinfo {author} {\bibfnamefont {Stephen~J}\
  \bibnamefont {Cotton}}, \bibinfo {author} {\bibfnamefont {Filip}\
  \bibnamefont {Wudarski}}, \bibinfo {author} {\bibfnamefont {Miroslav}\
  \bibnamefont {Urbanek}}, \bibinfo {author} {\bibfnamefont {Diptarka}\
  \bibnamefont {Hait}}, \bibinfo {author} {\bibfnamefont {Martin}\ \bibnamefont
  {Head-Gordon}}, \bibinfo {author} {\bibfnamefont {K~Birgitta}\ \bibnamefont
  {Whaley}}, \bibinfo {author} {\bibfnamefont {Jonathan}\ \bibnamefont
  {Moussa}}, \bibinfo {author} {\bibfnamefont {Nathan}\ \bibnamefont {Wiebe}},
  \emph {et~al.},\ }\bibfield  {title} {\enquote {\bibinfo {title} {Real time
  evolution for ultracompact hamiltonian eigenstates on quantum hardware},}\
  }\href@noop {} {\bibfield  {journal} {\bibinfo  {journal} {arXiv preprint
  arXiv:2103.08563}\ } (\bibinfo {year} {2021})}\BibitemShut {NoStop}%
\bibitem [{\citenamefont {Low}\ and\ \citenamefont
  {Chuang}(2019)}]{low2019hamiltonian}%
  \BibitemOpen
  \bibfield  {author} {\bibinfo {author} {\bibfnamefont {Guang~Hao}\
  \bibnamefont {Low}}\ and\ \bibinfo {author} {\bibfnamefont {Isaac~L}\
  \bibnamefont {Chuang}},\ }\bibfield  {title} {\enquote {\bibinfo {title}
  {Hamiltonian simulation by qubitization},}\ }\href@noop {} {\bibfield
  {journal} {\bibinfo  {journal} {Quantum}\ }\textbf {\bibinfo {volume} {3}},\
  \bibinfo {pages} {163} (\bibinfo {year} {2019})}\BibitemShut {NoStop}%
\bibitem [{\citenamefont {Poulin}\ \emph {et~al.}(2018)\citenamefont {Poulin},
  \citenamefont {Kitaev}, \citenamefont {Steiger}, \citenamefont {Hastings},\
  and\ \citenamefont {Troyer}}]{poulin2018quantum}%
  \BibitemOpen
  \bibfield  {author} {\bibinfo {author} {\bibfnamefont {David}\ \bibnamefont
  {Poulin}}, \bibinfo {author} {\bibfnamefont {Alexei}\ \bibnamefont {Kitaev}},
  \bibinfo {author} {\bibfnamefont {Damian~S}\ \bibnamefont {Steiger}},
  \bibinfo {author} {\bibfnamefont {Matthew~B}\ \bibnamefont {Hastings}}, \
  and\ \bibinfo {author} {\bibfnamefont {Matthias}\ \bibnamefont {Troyer}},\
  }\bibfield  {title} {\enquote {\bibinfo {title} {Quantum algorithm for
  spectral measurement with a lower gate count},}\ }\href@noop {} {\bibfield
  {journal} {\bibinfo  {journal} {Physical Review Letters}\ }\textbf {\bibinfo
  {volume} {121}},\ \bibinfo {pages} {010501} (\bibinfo {year}
  {2018})}\BibitemShut {NoStop}%
\bibitem [{\citenamefont {Gard}\ \emph {et~al.}(2020)\citenamefont {Gard},
  \citenamefont {Zhu}, \citenamefont {Barron}, \citenamefont {Mayhall},
  \citenamefont {Economou},\ and\ \citenamefont {Barnes}}]{gard2020efficient}%
  \BibitemOpen
  \bibfield  {author} {\bibinfo {author} {\bibfnamefont {Bryan~T}\ \bibnamefont
  {Gard}}, \bibinfo {author} {\bibfnamefont {Linghua}\ \bibnamefont {Zhu}},
  \bibinfo {author} {\bibfnamefont {George~S}\ \bibnamefont {Barron}}, \bibinfo
  {author} {\bibfnamefont {Nicholas~J}\ \bibnamefont {Mayhall}}, \bibinfo
  {author} {\bibfnamefont {Sophia~E}\ \bibnamefont {Economou}}, \ and\ \bibinfo
  {author} {\bibfnamefont {Edwin}\ \bibnamefont {Barnes}},\ }\bibfield  {title}
  {\enquote {\bibinfo {title} {Efficient symmetry-preserving state preparation
  circuits for the variational quantum eigensolver algorithm},}\ }\href@noop {}
  {\bibfield  {journal} {\bibinfo  {journal} {npj Quantum Information}\
  }\textbf {\bibinfo {volume} {6}},\ \bibinfo {pages} {1--9} (\bibinfo {year}
  {2020})}\BibitemShut {NoStop}%
\bibitem [{\citenamefont {Garcia-Escartin}\ and\ \citenamefont
  {Chamorro-Posada}(2013)}]{garcia2013swap}%
  \BibitemOpen
  \bibfield  {author} {\bibinfo {author} {\bibfnamefont {Juan~Carlos}\
  \bibnamefont {Garcia-Escartin}}\ and\ \bibinfo {author} {\bibfnamefont
  {Pedro}\ \bibnamefont {Chamorro-Posada}},\ }\bibfield  {title} {\enquote
  {\bibinfo {title} {Swap test and hong-ou-mandel effect are equivalent},}\
  }\href@noop {} {\bibfield  {journal} {\bibinfo  {journal} {Physical Review
  A}\ }\textbf {\bibinfo {volume} {87}},\ \bibinfo {pages} {052330} (\bibinfo
  {year} {2013})}\BibitemShut {NoStop}%
\bibitem [{\citenamefont {Cincio}\ \emph {et~al.}(2018)\citenamefont {Cincio},
  \citenamefont {Suba{\c{s}}{\i}}, \citenamefont {Sornborger},\ and\
  \citenamefont {Coles}}]{cincio2018learning}%
  \BibitemOpen
  \bibfield  {author} {\bibinfo {author} {\bibfnamefont {Lukasz}\ \bibnamefont
  {Cincio}}, \bibinfo {author} {\bibfnamefont {Yi{\u{g}}it}\ \bibnamefont
  {Suba{\c{s}}{\i}}}, \bibinfo {author} {\bibfnamefont {Andrew~T}\ \bibnamefont
  {Sornborger}}, \ and\ \bibinfo {author} {\bibfnamefont {Patrick~J}\
  \bibnamefont {Coles}},\ }\bibfield  {title} {\enquote {\bibinfo {title}
  {Learning the quantum algorithm for state overlap},}\ }\href@noop {}
  {\bibfield  {journal} {\bibinfo  {journal} {New Journal of Physics}\ }\textbf
  {\bibinfo {volume} {20}},\ \bibinfo {pages} {113022} (\bibinfo {year}
  {2018})}\BibitemShut {NoStop}%
\bibitem [{\citenamefont {Havl{\'\i}{\v{c}}ek}\ \emph
  {et~al.}(2019)\citenamefont {Havl{\'\i}{\v{c}}ek}, \citenamefont
  {C{\'o}rcoles}, \citenamefont {Temme}, \citenamefont {Harrow}, \citenamefont
  {Kandala}, \citenamefont {Chow},\ and\ \citenamefont
  {Gambetta}}]{havlivcek2019supervised}%
  \BibitemOpen
  \bibfield  {author} {\bibinfo {author} {\bibfnamefont {Vojt{\v{e}}ch}\
  \bibnamefont {Havl{\'\i}{\v{c}}ek}}, \bibinfo {author} {\bibfnamefont
  {Antonio~D}\ \bibnamefont {C{\'o}rcoles}}, \bibinfo {author} {\bibfnamefont
  {Kristan}\ \bibnamefont {Temme}}, \bibinfo {author} {\bibfnamefont {Aram~W}\
  \bibnamefont {Harrow}}, \bibinfo {author} {\bibfnamefont {Abhinav}\
  \bibnamefont {Kandala}}, \bibinfo {author} {\bibfnamefont {Jerry~M}\
  \bibnamefont {Chow}}, \ and\ \bibinfo {author} {\bibfnamefont {Jay~M}\
  \bibnamefont {Gambetta}},\ }\bibfield  {title} {\enquote {\bibinfo {title}
  {Supervised learning with quantum-enhanced feature spaces},}\ }\href@noop {}
  {\bibfield  {journal} {\bibinfo  {journal} {Nature}\ }\textbf {\bibinfo
  {volume} {567}},\ \bibinfo {pages} {209--212} (\bibinfo {year}
  {2019})}\BibitemShut {NoStop}%
\bibitem [{\citenamefont {Flammia}\ and\ \citenamefont
  {Liu}(2011)}]{flammia2011direct}%
  \BibitemOpen
  \bibfield  {author} {\bibinfo {author} {\bibfnamefont {Steven~T}\
  \bibnamefont {Flammia}}\ and\ \bibinfo {author} {\bibfnamefont {Yi-Kai}\
  \bibnamefont {Liu}},\ }\bibfield  {title} {\enquote {\bibinfo {title} {Direct
  fidelity estimation from few pauli measurements},}\ }\href@noop {} {\bibfield
   {journal} {\bibinfo  {journal} {Physical Review Letters}\ }\textbf {\bibinfo
  {volume} {106}},\ \bibinfo {pages} {230501} (\bibinfo {year}
  {2011})}\BibitemShut {NoStop}%
\bibitem [{\citenamefont {Elben}\ \emph {et~al.}(2019)\citenamefont {Elben},
  \citenamefont {Vermersch}, \citenamefont {Roos},\ and\ \citenamefont
  {Zoller}}]{elben2019statistical}%
  \BibitemOpen
  \bibfield  {author} {\bibinfo {author} {\bibfnamefont {Andreas}\ \bibnamefont
  {Elben}}, \bibinfo {author} {\bibfnamefont {Beno{\^\i}t}\ \bibnamefont
  {Vermersch}}, \bibinfo {author} {\bibfnamefont {Christian~F}\ \bibnamefont
  {Roos}}, \ and\ \bibinfo {author} {\bibfnamefont {Peter}\ \bibnamefont
  {Zoller}},\ }\bibfield  {title} {\enquote {\bibinfo {title} {Statistical
  correlations between locally randomized measurements: A toolbox for probing
  entanglement in many-body quantum states},}\ }\href@noop {} {\bibfield
  {journal} {\bibinfo  {journal} {Physical Review A}\ }\textbf {\bibinfo
  {volume} {99}},\ \bibinfo {pages} {052323} (\bibinfo {year}
  {2019})}\BibitemShut {NoStop}%
\bibitem [{\citenamefont {Elben}\ \emph {et~al.}(2020)\citenamefont {Elben},
  \citenamefont {Vermersch}, \citenamefont {van Bijnen}, \citenamefont
  {Kokail}, \citenamefont {Brydges}, \citenamefont {Maier}, \citenamefont
  {Joshi}, \citenamefont {Blatt}, \citenamefont {Roos},\ and\ \citenamefont
  {Zoller}}]{elben2020cross}%
  \BibitemOpen
  \bibfield  {author} {\bibinfo {author} {\bibfnamefont {Andreas}\ \bibnamefont
  {Elben}}, \bibinfo {author} {\bibfnamefont {Beno{\^\i}t}\ \bibnamefont
  {Vermersch}}, \bibinfo {author} {\bibfnamefont {Rick}\ \bibnamefont {van
  Bijnen}}, \bibinfo {author} {\bibfnamefont {Christian}\ \bibnamefont
  {Kokail}}, \bibinfo {author} {\bibfnamefont {Tiff}\ \bibnamefont {Brydges}},
  \bibinfo {author} {\bibfnamefont {Christine}\ \bibnamefont {Maier}}, \bibinfo
  {author} {\bibfnamefont {Manoj~K}\ \bibnamefont {Joshi}}, \bibinfo {author}
  {\bibfnamefont {Rainer}\ \bibnamefont {Blatt}}, \bibinfo {author}
  {\bibfnamefont {Christian~F}\ \bibnamefont {Roos}}, \ and\ \bibinfo {author}
  {\bibfnamefont {Peter}\ \bibnamefont {Zoller}},\ }\bibfield  {title}
  {\enquote {\bibinfo {title} {Cross-platform verification of intermediate
  scale quantum devices},}\ }\href@noop {} {\bibfield  {journal} {\bibinfo
  {journal} {Physical Review Letters}\ }\textbf {\bibinfo {volume} {124}},\
  \bibinfo {pages} {010504} (\bibinfo {year} {2020})}\BibitemShut {NoStop}%
\bibitem [{\citenamefont {Zhu}\ \emph {et~al.}(2021)\citenamefont {Zhu},
  \citenamefont {Cian}, \citenamefont {Noel}, \citenamefont {Risinger},
  \citenamefont {Biswas}, \citenamefont {Egan}, \citenamefont {Zhu},
  \citenamefont {Green}, \citenamefont {Maksymov}, \citenamefont {Nam} \emph
  {et~al.}}]{zhu2021cross}%
  \BibitemOpen
  \bibfield  {author} {\bibinfo {author} {\bibfnamefont {Daiwei}\ \bibnamefont
  {Zhu}}, \bibinfo {author} {\bibfnamefont {Ze-Pei}\ \bibnamefont {Cian}},
  \bibinfo {author} {\bibfnamefont {Crystal}\ \bibnamefont {Noel}}, \bibinfo
  {author} {\bibfnamefont {Andrew}\ \bibnamefont {Risinger}}, \bibinfo {author}
  {\bibfnamefont {Debopriyo}\ \bibnamefont {Biswas}}, \bibinfo {author}
  {\bibfnamefont {Laird}\ \bibnamefont {Egan}}, \bibinfo {author}
  {\bibfnamefont {Yingyue}\ \bibnamefont {Zhu}}, \bibinfo {author}
  {\bibfnamefont {Alaina~M}\ \bibnamefont {Green}}, \bibinfo {author}
  {\bibfnamefont {Andrii}\ \bibnamefont {Maksymov}}, \bibinfo {author}
  {\bibfnamefont {Yunseong}\ \bibnamefont {Nam}},  \emph {et~al.},\ }\bibfield
  {title} {\enquote {\bibinfo {title} {Cross-platform comparison of arbitrary
  quantum computations},}\ }\href@noop {} {\bibfield  {journal} {\bibinfo
  {journal} {arXiv preprint arXiv:2107.11387}\ } (\bibinfo {year}
  {2021})}\BibitemShut {NoStop}%
\bibitem [{\citenamefont {Huang}\ \emph {et~al.}(2020)\citenamefont {Huang},
  \citenamefont {Kueng},\ and\ \citenamefont {Preskill}}]{huang2020predicting}%
  \BibitemOpen
  \bibfield  {author} {\bibinfo {author} {\bibfnamefont {Hsin-Yuan}\
  \bibnamefont {Huang}}, \bibinfo {author} {\bibfnamefont {Richard}\
  \bibnamefont {Kueng}}, \ and\ \bibinfo {author} {\bibfnamefont {John}\
  \bibnamefont {Preskill}},\ }\bibfield  {title} {\enquote {\bibinfo {title}
  {Predicting many properties of a quantum system from very few
  measurements},}\ }\href@noop {} {\bibfield  {journal} {\bibinfo  {journal}
  {Nature Physics}\ }\textbf {\bibinfo {volume} {16}},\ \bibinfo {pages}
  {1050--1057} (\bibinfo {year} {2020})}\BibitemShut {NoStop}%
\bibitem [{\citenamefont {McArdle}\ \emph {et~al.}(2020)\citenamefont
  {McArdle}, \citenamefont {Endo}, \citenamefont {Aspuru-Guzik}, \citenamefont
  {Benjamin},\ and\ \citenamefont {Yuan}}]{alan_2020_review}%
  \BibitemOpen
  \bibfield  {author} {\bibinfo {author} {\bibfnamefont {Sam}\ \bibnamefont
  {McArdle}}, \bibinfo {author} {\bibfnamefont {Suguru}\ \bibnamefont {Endo}},
  \bibinfo {author} {\bibfnamefont {Al\'an}\ \bibnamefont {Aspuru-Guzik}},
  \bibinfo {author} {\bibfnamefont {Simon~C.}\ \bibnamefont {Benjamin}}, \ and\
  \bibinfo {author} {\bibfnamefont {Xiao}\ \bibnamefont {Yuan}},\ }\bibfield
  {title} {\enquote {\bibinfo {title} {Quantum computational chemistry},}\
  }\href {\doibase 10.1103/RevModPhys.92.015003} {\bibfield  {journal}
  {\bibinfo  {journal} {Rev. Mod. Phys.}\ }\textbf {\bibinfo {volume} {92}},\
  \bibinfo {pages} {015003} (\bibinfo {year} {2020})}\BibitemShut {NoStop}%
\bibitem [{\citenamefont {Bauer}\ \emph {et~al.}(2020)\citenamefont {Bauer},
  \citenamefont {Bravyi}, \citenamefont {Motta},\ and\ \citenamefont
  {Chan}}]{garnet_2020_review}%
  \BibitemOpen
  \bibfield  {author} {\bibinfo {author} {\bibfnamefont {Bela}\ \bibnamefont
  {Bauer}}, \bibinfo {author} {\bibfnamefont {Sergey}\ \bibnamefont {Bravyi}},
  \bibinfo {author} {\bibfnamefont {Mario}\ \bibnamefont {Motta}}, \ and\
  \bibinfo {author} {\bibfnamefont {Garnet Kin-Lic}\ \bibnamefont {Chan}},\
  }\bibfield  {title} {\enquote {\bibinfo {title} {Quantum algorithms for
  quantum chemistry and quantum materials science},}\ }\href {\doibase
  10.1021/acs.chemrev.9b00829} {\bibfield  {journal} {\bibinfo  {journal}
  {Chemical Reviews}\ }\textbf {\bibinfo {volume} {120}},\ \bibinfo {pages}
  {12685--12717} (\bibinfo {year} {2020})}\BibitemShut {NoStop}%
\bibitem [{\citenamefont {O'Brien}\ \emph {et~al.}(2019)\citenamefont
  {O'Brien}, \citenamefont {Tarasinski},\ and\ \citenamefont
  {Terhal}}]{O_Brien_2019}%
  \BibitemOpen
  \bibfield  {author} {\bibinfo {author} {\bibfnamefont {Thomas~E}\
  \bibnamefont {O'Brien}}, \bibinfo {author} {\bibfnamefont {Brian}\
  \bibnamefont {Tarasinski}}, \ and\ \bibinfo {author} {\bibfnamefont
  {Barbara~M}\ \bibnamefont {Terhal}},\ }\bibfield  {title} {\enquote {\bibinfo
  {title} {Quantum phase estimation of multiple eigenvalues for small-scale
  (noisy) experiments},}\ }\href {\doibase 10.1088/1367-2630/aafb8e} {\bibfield
   {journal} {\bibinfo  {journal} {New Journal of Physics}\ }\textbf {\bibinfo
  {volume} {21}},\ \bibinfo {pages} {023022} (\bibinfo {year}
  {2019})}\BibitemShut {NoStop}%
\bibitem [{\citenamefont {Somma}(2019)}]{Somma_2019}%
  \BibitemOpen
  \bibfield  {author} {\bibinfo {author} {\bibfnamefont {Rolando~D}\
  \bibnamefont {Somma}},\ }\bibfield  {title} {\enquote {\bibinfo {title}
  {Quantum eigenvalue estimation via time series analysis},}\ }\href {\doibase
  10.1088/1367-2630/ab5c60} {\bibfield  {journal} {\bibinfo  {journal} {New
  Journal of Physics}\ }\textbf {\bibinfo {volume} {21}},\ \bibinfo {pages}
  {123025} (\bibinfo {year} {2019})}\BibitemShut {NoStop}%
\bibitem [{\citenamefont {Sun}\ \emph {et~al.}(2018)\citenamefont {Sun},
  \citenamefont {Berkelbach}, \citenamefont {Blunt}, \citenamefont {Booth},
  \citenamefont {Guo}, \citenamefont {Li}, \citenamefont {Liu}, \citenamefont
  {McClain}, \citenamefont {Sayfutyarova}, \citenamefont {Sharma} \emph
  {et~al.}}]{sun2018pyscf}%
  \BibitemOpen
  \bibfield  {author} {\bibinfo {author} {\bibfnamefont {Qiming}\ \bibnamefont
  {Sun}}, \bibinfo {author} {\bibfnamefont {Timothy~C}\ \bibnamefont
  {Berkelbach}}, \bibinfo {author} {\bibfnamefont {Nick~S}\ \bibnamefont
  {Blunt}}, \bibinfo {author} {\bibfnamefont {George~H}\ \bibnamefont {Booth}},
  \bibinfo {author} {\bibfnamefont {Sheng}\ \bibnamefont {Guo}}, \bibinfo
  {author} {\bibfnamefont {Zhendong}\ \bibnamefont {Li}}, \bibinfo {author}
  {\bibfnamefont {Junzi}\ \bibnamefont {Liu}}, \bibinfo {author} {\bibfnamefont
  {James~D}\ \bibnamefont {McClain}}, \bibinfo {author} {\bibfnamefont
  {Elvira~R}\ \bibnamefont {Sayfutyarova}}, \bibinfo {author} {\bibfnamefont
  {Sandeep}\ \bibnamefont {Sharma}},  \emph {et~al.},\ }\bibfield  {title}
  {\enquote {\bibinfo {title} {{PySCF}: the python-based simulations of
  chemistry framework},}\ }\href@noop {} {\bibfield  {journal} {\bibinfo
  {journal} {Wiley Interdisciplinary Reviews: Computational Molecular Science}\
  }\textbf {\bibinfo {volume} {8}},\ \bibinfo {pages} {e1340} (\bibinfo {year}
  {2018})}\BibitemShut {NoStop}%
\bibitem [{\citenamefont {McClean}\ \emph {et~al.}(2020)\citenamefont
  {McClean}, \citenamefont {Rubin}, \citenamefont {Sung}, \citenamefont
  {Kivlichan}, \citenamefont {Bonet-Monroig}, \citenamefont {Cao},
  \citenamefont {Dai}, \citenamefont {Fried}, \citenamefont {Gidney},
  \citenamefont {Gimby} \emph {et~al.}}]{mcclean2020openfermion}%
  \BibitemOpen
  \bibfield  {author} {\bibinfo {author} {\bibfnamefont {Jarrod~R}\
  \bibnamefont {McClean}}, \bibinfo {author} {\bibfnamefont {Nicholas~C}\
  \bibnamefont {Rubin}}, \bibinfo {author} {\bibfnamefont {Kevin~J}\
  \bibnamefont {Sung}}, \bibinfo {author} {\bibfnamefont {Ian~D}\ \bibnamefont
  {Kivlichan}}, \bibinfo {author} {\bibfnamefont {Xavier}\ \bibnamefont
  {Bonet-Monroig}}, \bibinfo {author} {\bibfnamefont {Yudong}\ \bibnamefont
  {Cao}}, \bibinfo {author} {\bibfnamefont {Chengyu}\ \bibnamefont {Dai}},
  \bibinfo {author} {\bibfnamefont {E~Schuyler}\ \bibnamefont {Fried}},
  \bibinfo {author} {\bibfnamefont {Craig}\ \bibnamefont {Gidney}}, \bibinfo
  {author} {\bibfnamefont {Brendan}\ \bibnamefont {Gimby}},  \emph {et~al.},\
  }\bibfield  {title} {\enquote {\bibinfo {title} {{OpenFermion}: the
  electronic structure package for quantum computers},}\ }\href@noop {}
  {\bibfield  {journal} {\bibinfo  {journal} {Quantum Science and Technology}\
  }\textbf {\bibinfo {volume} {5}},\ \bibinfo {pages} {034014} (\bibinfo {year}
  {2020})}\BibitemShut {NoStop}%
\bibitem [{\citenamefont {Bergholm}\ \emph {et~al.}(2018)\citenamefont
  {Bergholm}, \citenamefont {Izaac}, \citenamefont {Schuld}, \citenamefont
  {Gogolin}, \citenamefont {Alam}, \citenamefont {Ahmed}, \citenamefont
  {Arrazola}, \citenamefont {Blank}, \citenamefont {Delgado}, \citenamefont
  {Jahangiri} \emph {et~al.}}]{bergholm2018pennylane}%
  \BibitemOpen
  \bibfield  {author} {\bibinfo {author} {\bibfnamefont {Ville}\ \bibnamefont
  {Bergholm}}, \bibinfo {author} {\bibfnamefont {Josh}\ \bibnamefont {Izaac}},
  \bibinfo {author} {\bibfnamefont {Maria}\ \bibnamefont {Schuld}}, \bibinfo
  {author} {\bibfnamefont {Christian}\ \bibnamefont {Gogolin}}, \bibinfo
  {author} {\bibfnamefont {M~Sohaib}\ \bibnamefont {Alam}}, \bibinfo {author}
  {\bibfnamefont {Shahnawaz}\ \bibnamefont {Ahmed}}, \bibinfo {author}
  {\bibfnamefont {Juan~Miguel}\ \bibnamefont {Arrazola}}, \bibinfo {author}
  {\bibfnamefont {Carsten}\ \bibnamefont {Blank}}, \bibinfo {author}
  {\bibfnamefont {Alain}\ \bibnamefont {Delgado}}, \bibinfo {author}
  {\bibfnamefont {Soran}\ \bibnamefont {Jahangiri}},  \emph {et~al.},\
  }\bibfield  {title} {\enquote {\bibinfo {title} {Pennylane: Automatic
  differentiation of hybrid quantum-classical computations},}\ }\href@noop {}
  {\bibfield  {journal} {\bibinfo  {journal} {arXiv preprint arXiv:1811.04968}\
  } (\bibinfo {year} {2018})}\BibitemShut {NoStop}%
\bibitem [{\citenamefont {Press}\ \emph {et~al.}(2007)\citenamefont {Press},
  \citenamefont {Teukolsky}, \citenamefont {Vetterling},\ and\ \citenamefont
  {Flannery}}]{Press2007}%
  \BibitemOpen
  \bibfield  {author} {\bibinfo {author} {\bibfnamefont {William~H.}\
  \bibnamefont {Press}}, \bibinfo {author} {\bibfnamefont {Saul~A.}\
  \bibnamefont {Teukolsky}}, \bibinfo {author} {\bibfnamefont {William~T.}\
  \bibnamefont {Vetterling}}, \ and\ \bibinfo {author} {\bibfnamefont
  {Brian~P.}\ \bibnamefont {Flannery}},\ }\href@noop {} {\emph {\bibinfo
  {title} {Numerical Recipes 3rd Edition: The Art of Scientific Computing}}},\
  \bibinfo {edition} {3rd}\ ed.\ (\bibinfo  {publisher} {Cambridge University
  Press},\ \bibinfo {address} {Cambridge},\ \bibinfo {year} {2007})\BibitemShut
  {NoStop}%
\end{thebibliography}%

\newpage

\newpage

\newpage
\section*{Appendix A: Quantum chemistry and molecular systems studied}

Within the Born-Oppenheimer approximation, a molecule is comprised of $\eta$ electrons interacting within a potential produced by nuclei at fixed positions. Using the second quantization formalism, the problem may be cast in terms of $N$ single-particle spin orbitals that can be occupied or empty. In the absence of external fields, the non-relativistic molecular Hamiltonian is written as:
\begin{equation}
    H = h_{\text{nuc}} + \sum_{pq} h_{pq} a_p^\dagger a_q + \tfrac{1}{2}\sum_{pqrs} h_{pqrs} a^\dagger_p a^\dagger_q a_r a_s
\end{equation}
where $a_p$ and $a_p^\dagger$ correspond to fermionic annihilation and creation operators that obey the anti-commutation relations, $\{a_j,a_k \}=0$, $\{a_j^\dagger,a_k^\dagger \}=0$, and $\{a_j,a_k^\dagger \}=\delta_{jk}$. $h_{\text{nuc}}$ corresponds to the classical electrostatic repulsion between nuclei, while $ h_{pq}$ and $h_{pqrs}$ correspond to one- and two-electron integrals and written explicitly as:
\begin{align}
   h_{\text{nuc}} &= \frac{1}{2}\sum_{i\neq j} \frac{Z_iZ_j}{| 
   \mathbf{R}_i - \mathbf{R}_j|}, \\
   h_{pq} &= \int\! d\sigma \; \phi_p^*(\sigma)\left( -\frac{\nabla_r^2}{2} - \sum_i \frac{Z_i}{|\mathbf{R}_i - \mathbf{r}|} \right) \phi_q(\sigma), \\
   h_{pqrs} &= \int \! d\sigma_1 d\sigma_2 \;\; \frac{\phi_p^*(\sigma_1)\phi_q^*(\sigma_2)\phi_r(\sigma_1)\phi_s(\sigma_2)}{| 
   \mathbf{r}_i - \mathbf{r}_j|}.
\end{align}
where the summations run over all nuclei. $Z_i$ represents the nuclear charge, $\mathbf{r}$ and $\mathbf{R}$ denote electronic and nuclear spatial coordinates, and $\sigma$ is a generalized coordinate consisting of the spatial and spin degrees of freedom, $\sigma_i = (\mathbf{r}_i,s_i)$. The function $\phi(\sigma)$ represents a one-electron spin-orbital. In the main text, we calculate the potential energy surface $E(\mathbf{R})$, which can be used to describe a wide range of processes such as reaction dynamics, bond-breaking, and chemical dynamics. In order to predict thermochemical properties accurately at room temperature and atmospheric pressure, we require an estimate of the potential energy surface within a chemical accuracy of $1$ kcal/mol or equivalently $1.59\times 10^{-3}$ Hartrees or 43.4 meV, represented by the horizontal lines in the energy error plots in the manuscript. 

For our illustrative calculations, we consider (i) a 6-atom linear hydrogen chain, H$_6$, (ii) the bent H$_2$O molecule, and (iii) the linear BeH$_2$ molecule. We employ minimal STO-3G bases such that there are six spatial molecular orbitals for H$_6$, and seven molecular orbitals each for H$_2$O and BeH$_2$. The number of qubits, after performing the Jordan-Wigner transformation is equal to the number of (active) spin-orbitals being treated. The hydrogen chain can thus be treated with twelve qubits corresponding to the total number of spin-orbitals and within full configuration interaction (FCI). For H$_2$O and BeH$_2$ we employ a complete active space approach wherein there are $n$ active electrons distributed in $m$ spatial orbitals with a full configuration interaction being carried out only among the active electrons and orbitals, sometimes denoted as CASCI$(n,m)$.  Generally the $m$ orbitals are the highest energy
ones and the remaining electrons are effectively frozen in the lower, inactive orbitals.
In the case of H$_2$O we use CASCI(8,6) and for BeH$_2$ we use CASCI(6,6), both corresponding to twelve qubit Hamiltonians in both cases.  Of course the levels of theory here, in particular the use of STO-3G basis sets,  is quite crude by computational chemistry standards, but it is sufficient to illustrate and compare the various diagonalization procedures of interest to us.

To obtain the coefficients of the second quantized Hamiltonian, we defined the structure of the molecule by its constituent elements and nuclear coordinates of the atoms. The initial basis of single-particle states was obtained via the Hartree-Fock method where each electron is treated as an independent particle that moves under the influence of the Coulomb potential due to the nuclei as well as a mean field generated by all of the other electrons. We used the quantum chemistry package PySCF \cite{sun2018pyscf} to obtain the optimized coefficients of the linear combination of the atomic orbitals, obtained via the Hartree-Fock method. To map the fermionic second quantization Hamiltonian to a qubit-basis, it is possible to use the Jordan-Wigner, Parity, or Bravyi-Kitaev transformations. We used the OpenFermion Python package \cite{mcclean2020openfermion} to obtain the qubit representation of the Hamiltonian, where the Jordan-Wigner basis was chosen for all of the numerical simulations. All of these packages can also be found within the Pennylane cross-platform Python library \cite{bergholm2018pennylane}.
\vspace{0.2cm}

\section*{Appendix B: Choice of time step}

First we consider the case that the generalized eigenvalue problem involves
$f(\hat{H})$ = $\hat{H}$. If an absolute energy scale is used for the
corresponding Hamiltonian matrix such that the energy  eigenvalues of interest
could be large in magnitude, the time step $\tau$ should ideally be equal to or smaller
than  $\pi/\Delta E$ where $\Delta E$ is an upper bound to
the spectral range (difference between maximum and minimum eigenvalues) and atomic
units ($\hbar$ = 1) are assumed.
This parallels
considerations of discrete Fourier transforms wherein the possible range of
eigenvalues is determined from the Nyquist interval (in terms of angular
frequencies or energies), $-\pi/\tau$ to $+\pi/\tau$ \cite{Press2007}.   Note that it is generally not difficult to make such  an upper
limit estimate to the spectral range given some knowledge of the problem,
e.g., the Hartree-Fock and molecular orbital energies in the case of the
electronic structure problem.

If sufficient accuracy can still be achieved with the corresponding
Trotter time steps, however, one can actually use
larger $\tau$ values than $\pi /\Delta E$ if the energy zero of
the Hamiltonian is simply shifted to be in the center of the
desired energy range, i.e. the Hamiltonian in all equations is taken to
be the energy-shifted one.  Aliasing, i.e., mapping of eigenvalues outside
the corresponding Nyquist interval into it can be easily and cheaply
monitored for:  a slightly different $\tau$ will yield the same ``true’’
eigenvalues but the aliased ones will shift.  Actually, in the three
chemical systems considered in this paper there were no issues with
aliasing at all because even with the energy shifting permitting the
larger $\tau$ values, the Nyquist interval was still tens of
Hartrees wide and sufficient to capture all relevant eigenvalues present.

Similar considerations to the above apply when the generalized eigenvalue
problem involves $f(\hat{H})$ = $\exp (-i\hat{H}\tau)$.  That is,
ideally $\tau$ $\le$ $\pi/\Delta E$ where $\Delta E$ is an upper bound to the
Hamiltonian's spectral range but, with appropriate recognition of
aliasing issues larger $\tau$ could be used.  We note that in this case
the eigenvalues $f(E_k)$ are complex and lie on the unit circle.
Let $\theta_k$ = $\tan^{-1} (-\rm{Im}~ f(E_k)/\rm{Re}~ f(E_k))$ be
the  angle
between $-\pi$ and $\pi$  that results from typical numerical arctangent
function (``$\rm{atan2(y, x)}$’’) calls.  The possible energy eigenvalues are
$E_k$ = $(\theta_k + j 2\pi)/\tau$, where $j$ = 0, $\pm 1$,  $\pm 2$,
etc.
If some absolute energy scale that leads to large magnitude physical energy
eigenvalues is used, then $|j|$ $\ge$ 1 value may have to be used to
obtain energy eigenvalues in the desired physical range.
The Hamiltonian energy shifting noted above is particularly
convenient for this case because then, typically, it is not necessary
to shift the eigenvalues and $j$ = 0 suffices.
\vspace{0.2cm}

\section*{Appendix C: Numerical solution of the generalized eigenvalue problem}

The numerical simulations shown in Figures 2 to 4 were performed with an in-house Python code using standard $\texttt{NumPy}$ and $\texttt{SciPy}$ numerical linear algebra packages. The classical solution of the complex generalized eigenvalue problem, $\bf{F} \bf{c}$. = $f \bf{S} \bf{c}, $ was performed using the QZ algorithm or generalized Schur decomposition~\cite{GoluVanl96} of the complex matrices $\bf{F}$ and $\bf{S}$, which we found to be much more stable compared to conventional generalized eigenvalue solvers. 

An alternative, more ``hands on'' approach to solving the generalized eigenvalue problem is to carry out singular value decomposition ~\cite{GoluVanl96} of the overlap matrix, $\bf{S}$, thereby allowing construction of its inverse and turning the problem into an eigenvalue problem for a complex matrix of the form, e.g.,  $\bf{S}^{-1}\bf{F} \bf{c} $ = $f \bf{c}$. The latter problem can be numerically solved with standard numerical software but may require appropriate zeroing of some of the singular values when the inverse is constructed.

\vspace{0.2cm}

\section*{Appendix D: Overview of Hadamard Test }

In the following, we describe the Hadamard test used for estimating the matrix element, $\braket{\phi_i|U|\phi_i}$. This method uses an ancilla qubit initially prepared in the $\ket{0}$ state, with the rest of the quantum circuit shown below:
\[
\Qcircuit @C=1em @R=1em {
\lstick{\ket{0}} & \gate{H} & \ctrl{1}&  \meter  \qw \\
\lstick{\ket{\phi_i}} & \qw {/}  & \gate{U} & \qw 
}
\]
This type of approach will effectively require an ancilla with one-to-all connectivity in order to keep the circuit gate depth as short as possible. The real and imaginary parts of the matrix element are obtained by measuring the ancilla qubit in the X and Y Pauli basis, where it can be shown that $\braket{\sigma_x+i\sigma_y} = \braket{\phi_i|U|\phi_i}$. This type of Hadamard quantum circuit is sufficient for single-reference-based Krylov subspace algorithms. For more general off-diagonal matrix elements of the form, $\braket{\phi_i|\phi_j} = \braket{0|U_i^\dagger U_j|0}$, as required by multi-reference Krylov subspace algorithms \cite{stair2020multireference}, the following quantum circuit would be used:
\[
\Qcircuit @C=1em @R=1em @!R {
\lstick{\ket{0}\;\;\;} & \gate{H} & \gate{X} & \ctrl{1}& \gate{X} & \ctrl{1} & \meter  \qw \\
\lstick{\ket{0}^{\small{\otimes N}} \!\!\!} & \qw {/} & \qw & \gate{U_j} & \qw &   \gate{U_i} & \qw
}
\]
where $U_i$ and $U_j$ correspond to the quantum circuits that prepare $\ket{\phi_i}$ and $\ket{\phi_j}$, defined through the relations $\ket{\phi_i} = U_i\ket{0}$ and $\ket{\phi_j} = U_j\ket{0}$ respectively. The real and imaginary parts are obtained through the measurement of the ancilla qubit in the $X$ and $Y$ Pauli basis as before.

\section*{Appendix E: Details of excited-state energy calculations}

In the main manuscript, we performed excited-state energy calculations for a water molecule as a function of the bond length $R$ between the oxygen and hydrogen atoms. We only considered the singlet excited-state energies where the total angular momentum is zero. In principle, it is possible to use the Hartree-Fock state as the initial starting state $\ket{\phi_o}$ since it corresponds to a singlet state and, by symmetry, it will be connected to the excited singlet states. However, the convergence towards the excited-state energies will be slow because the Hartree-Fock state is nearly orthogonal to the desired excited-state wavefunctions. Here, we propose the use of physically-motivated ansatz states to find the first four singlet excited-state energies of water. The ansatze consists of the following four states:
\begin{align}
    \ket{\Phi_0} &= \tfrac{1}{\sqrt{2}}\left(\ket{0 0 0 1 1 0 1 1 1 1 1 1} - \ket{0 0 1 0 0 1 1 1 1 1 1 1} \right) \\
    \ket{\Phi_1} &= \tfrac{1}{\sqrt{2}}\left(\ket{0 0 0 1 1 1 1 0 1 1 1 1} - \ket{0 0 1 0 1 1 0 1 1 1 1 1} \right) \\
    \ket{\Phi_2} &= \tfrac{1}{\sqrt{2}}\left(\ket{0 1 0 0 1 0 1 1 1 1 1 1} - \ket{1 0 0 0 0 1 1 1 1 1 1 1} \right) \\
    \ket{\Phi_3} &= \tfrac{1}{\sqrt{2}}\left(\ket{0 1 0 0 1 1 1 0 1 1 1 1} - \ket{1 0 0 0 1 1 0 1 1 1 1 1} \right)
\end{align}
where we use an alternating alpha (spin up) beta (spin down) ordering for the spin-orbitals with increasing energies from right to left. Note that each of these states has zero total angular momentum, therefore, they will be decoupled from any triplet states that might otherwise slow down the convergence of the Krylov-based algorithm. For the calculations, we used states (18)-(21) as the initial single-reference starting state $\ket{\phi_o}$. Each of the calculations are independent from one another. To perform the multi-fidelity estimation protocol from the main manuscript, each of these four states would require a state preparation step that initializes the superposition state, $\tfrac{1}{\sqrt{2}}(\ket{0}^{\otimes{N}} + \ket{\Phi_k})$, where $k=0,1,2,3$, on the quantum computer. 


\end{document}